\documentclass[onecolumn,linenumbers]{revtex4}
\usepackage{graphicx}
\usepackage{epstopdf}
\usepackage{caption}
\usepackage{graphicx}
\usepackage{bm}
\usepackage[latin1]{inputenc}
\usepackage{amsmath}
\usepackage{amsfonts}
\usepackage{subfigure}
\usepackage{amssymb}
\usepackage{makeidx}
\usepackage{color}
\usepackage{mathrsfs}
\RequirePackage[colorlinks,citecolor=blue,urlcolor=blue,linkcolor=magenta]{hyperref}

\newcommand {\drm} {\dot{\rho}_{\rm mat}}
\newcommand {\rhm} {\rho_{\rm mat}}

\newcommand {\Om} {\Omega_{\rm mat}}
\begin{document}

\title{Cosmological dynamics of brane gravity: A global dynamical system perspective}

\smallskip

\author{\bf~ Hmar~Zonunmawia$^1$\footnote{zonunmawiah@gmail.com}, Wompherdeiki Khyllep$^{1,2}$\footnote{sjwomkhyllep@gmail.com},
  Jibitesh Dutta$^{3,4}$\footnote{jdutta29@gmail.com,~jibitesh@nehu.ac.in} and Laur J\"arv $^5$ \footnote{laur.jarv@ut.ee}}
\smallskip

\affiliation{$^1$Department of Mathematics,~ North Eastern Hill
University,~NEHU Campus,  Shillong, Meghalaya 793022, India}

\affiliation{$^{2}$ Department of Mathematics, St. Anthony's College, Shillong, Meghalaya 793001, India}

\affiliation{$^{3}$ Mathematics Division, Department of Basic
Sciences and Social Sciences,~ North Eastern Hill University,~NEHU
Campus, Shillong, Meghalaya 793022, India}

\affiliation{$^4$ Inter University Center for Astronomy and Astrophysics, Pune 411 007, India}

\affiliation{$^5$ Laboratory of Theoretical Physics, Institute of Physics, University of Tartu,
W. Ostwaldi 1, 50411 Tartu, Estonia}

%\date{\today}
\begin{abstract}
The braneworld model of gravity is well-known for several notable cosmological features such as self-acceleration originating
from a geometric and not matter source, effective dark energy behavior with phantom characteristics but not leading to a Big-Rip singularity, rough resemblance
to the $\Lambda$CDM evolution, etc. The dynamical system tools usually allow us to obtain generic conclusions on the global dynamics of a system over a wide range of initial conditions.
 With this motivation, in order to recover the important features of the braneworld model from a more global perspective, here, we investigate the global cosmological dynamics of the
 braneworld model using dynamical system techniques. We  first analyze the case where there is just a normal matter on the brane and then extend the analysis to the case
  with an extra scalar field also trapped on the brane. In the presence of a scalar field, potentials belonging to different classes are considered.
  The stability behavior of critical points is examined using linear stability analysis and when necessary center manifold theory as well as numerical perturbation techniques are also used. To understand the global dynamics of a dynamical system, we utilized the Poincar\'e compactification method to capture the properties of all possible critical points. Applying dynamical system analysis, we found that brane gravity is consistent with observed actions of the Universe. In particular, our analysis shows that important cosmological behaviors like the long-lasting matter-dominated era, late time acceleration as well as the avoidance of Big-Rip singularity can be realized in brane gravity for a wide range of initial conditions.

\end{abstract}

 \maketitle

\section{Introduction}
The accelerated expansion of the Universe as confirmed by various surveys and experiments \cite{Ade:2013zuv,Ade:2015xua} has led to various possible theoretical explanations of this phenomenon. One approach is the introduction of an extra quantity called dark energy (DE) as the matter component of the Einstein field equation. The cosmological constant $\Lambda$ with Lorentz invariant equation of state is the simplest and successful choice of DE in spite of some theoretical issues associated with it \cite{Martin:2012bt,Caldwell:1997ii}. Another frequently used form of DE is the dynamical scalar field which reproduces the cosmological constant behavior at late times. Scalar field models are well motivated by the low-energy limit of some well-known high-energy theories like the string theory. They could also play an important role in solving problems associated with inflation, dark matter besides DE. For a review on different cosmological DE models see \cite{Copeland:2006wr,Tsujikawa:2013fta}. The story so far on various aspects of DE candidates is summarized in an excellent  recent work by Brax \cite{brax}. Another intriguing paradigm to explain the accelerated expansion of the Universe is by modifying the geometrical side of the Einstein field equations. This opens the way for a plethora of modified gravity theories which are theoretically and observationally interesting for sensitive and precise surveys (see \cite{clifton,Koyama:2015vza,Nojiri:2017ncd} for reviews).

One of the well developed and extensively studied modified gravity
settings, motivated by the superstring and M-theory is the concept
of an extra dimension known as \textit{braneworld}. In a braneworld,
the observable universe is a lower dimensional hyperspace known as
the `brane' is embedded in a higher dimensional space called the
`bulk'. While all the standard model fields are assumed to be
trapped on the brane, only gravity can propagate to the bulk. In
this theory, the bulk dimension orthogonal to the brane is not
compact and could even be of infinite length. This makes it
different from the Kaluza-Klein construction. Braneworld theories
usually lead to notable cosmological consequences. For example, one
of the popular braneworld models is the Randall-Sundrum (RS) model
\cite{Randall:1999ee,Randall:1999vf}, which modifies gravity at
small scales and has an impact on the early inflationary universe.
Interestingly, in the RS model, the inflaton field may survive till
the present universe and plays the role of quintessence field
\cite{Majumdar:2001mm, Sahni:2001qp}. Another well-known braneworld
model is the Dvali-Gabadadze-Porrati (DGP) model which modifies
gravity at large scales and has an impact on the late universe
\cite{Dvali:2000hr,Dvali:2000xg,Collins:2000yb}. In this model, the
accelerated expansion of the late universe can be explained as a
result of the leakage of gravity into the bulk without introducing
DE \cite{Deffayet:2000uy,Deffayet:2001pu}.

A general brane gravity based model was introduced by Sahni and Shtanov \cite{Sahni:2002dx} and discussed in detail in
Refs.\ \cite{Sahni:2004fb,Viznyuk:2013ywa,Bag:2016tvc,Viznyuk:2018eiz}. In specific limits, this model reduces to RS or DGP models, or just to
the case of general relativity (GR).  The main motivation of this model is the existence of cosmological solutions which are phantom-like but do
 not lead to any instabilities such as Big-Rip singularity. It is worth mentioning that phantom DE is slightly favored by some recent
 observations \cite{Ade:2013zuv,Ade:2015xua,Rest:2013mwz,Delubac:2014aqe} but the standard phantom scalar field is plagued by such instabilities.
  Apart from the phantom behavior, this braneworld model also leads to other cosmological surprises like the early and late time
  unification (Quintessential Inflation), transient acceleration, quiescent future singularity, cosmological loitering, cosmic mimicry property
  (mimicking the $\Lambda$CDM model at late times); for a brief summary  of important features of this braneworld model see \cite{Sahni:2005pf,Sahni:2008zb}. This braneworld model has been also used to address galactic rotational curves as an alternative explanation to dark matter \cite{Viznyuk:2007ft}. The analysis of this braneworld model at the perturbation level beyond the quasi-static approximation shows a significant difference from the $\Lambda$CDM model \cite{Viznyuk:2013ywa,Bag:2016tvc,Shtanov:2007dh,Bag:2018jle}. Interestingly, the expansion rate of the Phantom brane is slower than the $\Lambda$CDM which allows the braneworld to fit with measurements of Hubble rate reported in \cite{Delubac:2014aqe,Bautista:2017zgn,Zhao:2018jxv}. Further, the stability analysis of large-scale structures within this braneworld framework along with the presence of bulk cosmological constant was performed in \cite{Bhattacharya:2017ydy}. A confrontation of this model with various distance measures from SNe type-Ia and BAO observations, as well as compressed CMB data to constrain model parameters have been performed in \cite{Alam:2016wpf}. These distance measures signal the consistency of extra dimensions with present astronomical probes.

Dynamical systems techniques are useful tools to study the complete qualitative dynamical behavior of a cosmological model, without analytically solving the highly non-linear set of differential equations. In this approach, one has to extract all possible critical points lying at the finite region of the phase space as well near infinity (if the phase space is not compact). To capture possible critical points near infinity, the Poincar\'e compactification method which maps all points near infinity to points on the boundary of the Poincar\'e sphere is usually used \cite{slynch}.  For some recent work where these methods of dynamical systems have been used, a reader can refer to Refs.\ \cite{Leon:2014rra,Tamanini:2014nvd,Dutta:2016bbs,Dutta:2017kch}. The story so far on the applications of dynamical systems in DE and modified gravity models have been summarized in a recent comprehensive review by Bahamonde \textit{et al.} \cite{Bahamonde:2017ize}.

Dynamical evolution of different braneworld models like the RS model and DGP model has been studied using the dynamical systems tools before. For instance, the analysis of the RS model in Friedmann Robertson Walker (FRW) and Bianchi type background was analyzed in Refs.~\cite{Campos:2001pa,Campos:2001cn}. Various papers have studied the cosmological dynamics of RS and DGP model in the presence of scalar field with an exponential potential \cite{Quiros:2008hv,Nozari,Biswas:2015zka}. These works have been extended to various classes of potentials \cite{Leyva:2009zz,Gonzalez:2008wa,Dutta:2016dnt}. Dynamical systems analysis for the general braneworld model have been performed in the absence of scalar field in \cite{Iakubovskyi:2004ze} but no such study has been conducted in the presence of an extra scalar field. Moreover, in all previous work, a global analysis of the stability of critical points at finite as well as the infinite regime has not been carried out using the more advanced tools of dynamical systems.

In this work, we shall perform the global analysis of cosmological dynamics of the generic braneworld model using dynamical systems methods.
 We shall first analyze the case where there is just a normal matter on the brane, and then extend the analysis to the case with an extra scalar field also
 trapped on the brane. In the presence of scalar field, we consider only a class of scalar field potentials, $V(\phi)$,
 where the potential parameter $\Gamma \Big(=V \frac{d^2V}{d \phi^2}\left(\frac{dV}{d \phi}\right)^{-2}\Big)$ can be written explicitly as a function
 of  another potential parameter $s(=-\frac{1}{V}\frac{dV}{d \phi})$. This method has been quite successful in determining
 the dynamics of cosmological models for general potentials (for some recent work see Refs.\ \cite{Dutta:2016dnt,Dutta:2016bbs,Dutta:2017kch,Zonunmawia:2017ofc}).
  Further, in order to comprehend the cosmological dynamics, we consider three concrete potentials as examples.
  It is observed that the potential plays a major role in the late time behavior of the universe. It is worth mentioning
   that the results obtained in the present work match with the ones reported in previous literature. However, by employing
   the dynamical systems tools in the present work gives a general conclusion on the cosmological dynamics for a generic choice of initial conditions.
   Thus, the main scope of the present work is to provide a preliminary test of this braneworld model for further investigation.

The organization of the paper is as follows: In Sec. \ref{sec:cosmological_eqn}, we briefly review the basic equations of the braneworld model introduced in \cite{Sahni:2002dx} and also form the autonomous system of differential equations for a spatially flat homogeneous and isotropic universe. In Sec. \ref{sec:stability_pts_no_scalar}, we first analyze the cosmological behavior of the braneworld model without any scalar field.
In Sec. \ref{sec:stability_pts}, we determine the stability and cosmological behavior of critical points for three types of  potentials: \ref{subsec:exp} for potential with $\Gamma(s)=1$ for all $s$, \ref{subsec:cosh} for potential with $\Gamma(s)=1$ for some $s$ and \ref{subsec:pow} for potential with $\Gamma(s)\neq 1$ for any $s$.   The cosmological implications of the braneworld model are highlighted in Sec. \ref{sec:cos_imp}. Finally, the conclusion is given in Sec. \ref{sec:conc}.

\section{Generic braneworld model} \label{sec:cosmological_eqn}
The total action of the braneworld model is given by  \cite{Sahni:2002dx}
%%%%%%%%%%%%%%%%%%%%%%%%%%%%%%%%%%
\begin{equation}\label{1}
S= M^3 \Bigg[\int_{bulk} (R_5-2\Lambda)-2 \int_{brane} K\Bigg]+\int_{brane}(m^2 R_4-2 \sigma)+\int_{brane}L(h_{\mu\nu},\psi),
\end{equation}
%%%%%%%%%%%%%%%%%%%%%%%%%%%%%%%%%%%%%%%%%%%%%%%%%%%%%%%%%%%%%%%%%%%%%%%%%%%
where $R_5$ is the scalar curvature corresponds to the metric $g_{ab}$ ($a,b=0,1,2,3,4$) of the five-dimensional bulk, and $R_4$ is the scalar curvature corresponding to the induced metric $h_{\mu\nu}$ ($\mu, \nu=0,1,2,3$) on the brane. The symbol $L(h_{\mu\nu},\psi)$ denotes the Lagrangian density of the four-dimensional matter field $\psi$ which is restricted to the brane and interact only with the metric $h_{\mu\nu}$. The scalar $K=K_{\mu\nu}h^{\mu\nu}$ is the trace of the symmetric tensor of the extrinsic curvature $K_{\mu\nu}=h^\alpha_\mu \nabla_\alpha n_\nu$ with respect to  the vector field of the outer unit normal to the brane $n^{\mu}$. In the above action,  integrations over the bulk and brane are respectively taken with respect to the natural volume  elements $\sqrt{-g}\,{\rm d}^5x$ and  $\sqrt{-h}\,{\rm d}^4x$ where $g$ and $h$ denote the determinants corresponding to metrics $g_{ab}$ and $h_{\mu\nu}$. The quantities $M$ and $m$ denote the five-dimensional and four-dimensional Planck masses, while the quantities $\Lambda$ and $\sigma$ stand for the bulk cosmological constant and the brane tension, respectively.

The generic action \eqref{1} under certain conditions reduces to the following important sub-classes:

\begin{itemize}
\item The RS model, by setting the limiting condition  $m \rightarrow 0$ in the action \eqref{1};
\item The DGP model by setting both the bulk cosmological constant and brane tension to zero, i.e.\ $\Lambda=0$ and $\sigma=0$ in the action \eqref{1};
\item Lastly, it reduces to GR by setting $M=0$ in the action \eqref{1} with $1/m^2$ playing the role of the gravitational constant.
\end{itemize}

The action \eqref{1} gives rise to the following Einstein field equation in the bulk \cite{Sahni:2005mc,Shiromizu:1999wj}
\begin{equation}
G_{ab}+\Lambda\,g_{ab}=0,
\end{equation}

%%%%%%%%%%%%%%%%%%%%%%%%%%%%%%%%%%%%%%%%%%%%%%%%%%%%%%%%%%%%%%%%%%%%%%%%%%%%%%%%%%%%%%%%%%%%%%%%%%%%%%%%%

and the following effective equation on the brane
  \begin{equation}\label{EFE}
 G_{\mu\nu}+\Big(\frac{\Lambda_{RS}}{b+1}\Big)h_{\mu\nu}=\Big(\frac{b}{b+1}\Big)\frac{1}{m^2}T_{\mu\nu}+\Big(\frac{1}{b+1}\Big)\Bigg[\frac{1}{M^6}Q_{\mu\nu}-C_{\mu\nu}\Bigg],
 \end{equation}
 %%%%%%%%%%%%%%%%%%%%%%%%%%%%%%%%%%%%%%%%%%%%%%%%%%%%%
 where
 \begin{equation}\label{EFE1}
 b=\frac{\sigma l}{3m^3}, \qquad l=\frac{2m^2}{M^3}, \qquad \Lambda_{RS}=\frac{\Lambda}{2}+\frac{\sigma^2}{3M^6},
 \end{equation}
  %%%%%%%%%%%%%%%%%%%%%%%%%%%%%%%%%%%%%%%%%%%%%%%%%%%%%%%%%%%%
    \begin{equation} \label{EFE2}
  Q_{\mu\nu}=\frac{1}{3}EE_{\mu\nu}-E_{\mu\lambda}E^\lambda_\nu+\frac{1}{2}\Big(E_{\rho\lambda}E^{\rho\lambda}-\frac{1}{3}E^2\Big)h_{\mu\nu},
    \end{equation}
%%%%%%%%%%%%%%%%%%%%%%%%%%%%%%%%%%%%
  \begin{equation} \label{EFE3}
  E_{\mu\nu}= m^2G_{\mu\nu}-T_{\mu\nu}, \hspace{0.2cm} E=E^\mu_\mu.
  \end{equation}
 %%%%%%%%%%%%%%%%%%%%%%%%%%%%%%%%%%%%%%%%%%%%%%%%%%%%%%%%%%%%%%%%%%%%%%%%%%%%%%%%
Due to the presence of the symmetric traceless tensor $C_{\mu\nu}$ in \eqref{EFE} (arises from the projection of the five-dimensional Weyl tensor from the bulk onto the brane) the dynamics on the brane is not closed. The tensor  $C_{\mu\nu}$ on the brane is related to the tensor $Q_{\mu\nu}$ via the conservation equation
 %%%%%%%%%%%%%%%%%%%%%%%%%%%%%%%%%%%%%%%%%%%%%%%%%%%%%%%%
  \begin{equation}\label{EFE4}
  \nabla^\mu(Q_{\mu\nu}-M^6\,C_{\mu\nu})=0,
  \end{equation}
which results from the application of  Bianchi identity to \eqref{EFE} and the law of the conservation of the stress energy tensor of matter
  \begin{align}
  \nabla^\mu T_{\mu \nu}=0.
  \end{align}
  %%%%%%%%%%%%%%%%%%%%%%%%%%%%%%%%%%%%%%%%%%%%%%%%%%%%%%%%%%%%%%%
Here, $\nabla^\mu$ denotes the covariant derivative on the brane with respect to the induced metric $h_{\mu \nu}$.

To be in line with current observations, the brane is assumed to be homogeneous and isotropic characterized by Friedmann-Robertson-Walker (FRW) metric
\begin{equation}\label{metric}
ds^2=-dt^2+a^2(t)\Bigg[\frac{dr^2}{1-kr^2}+r^2\left(d\theta^2+\sin^2 \theta d\phi^2\right)\Bigg],
\end{equation}
where $k=-1, 0$ or $+1$ while $a(t)$ is the scale factor and $t$ is the cosmological time. Using the metric \eqref{metric}, one can obtain from Eq.\ \eqref{EFE}, the following modified Friedmann equation \cite{Sahni:2002dx,Deffayet:2001pu}
\begin{equation}\label{friedman0}
H^2+\frac{k}{a^2}=\frac{\rho+\sigma}{3m^2}+\frac{2}{l^2}\Bigg[1\pm\sqrt{1+l^2\Big(\frac{\rho+\sigma}{3m^2}-\frac{\Lambda}{6}-\frac{C}{a^4}\Big)}\Bigg],
\end{equation}
where $H=\frac{\dot a}{a}$ is the Hubble parameter, {\bf $l$ defined in \eqref{EFE1} is the crossover length characterizing the transition between 4 dimensional and 5 dimensional gravities}, $\rho$ is the total energy density of the matter fields on the brane and $C$ is the constant corresponding to the symmetric traceless tensor $C_{\mu\nu}$ which stands for the dark radiation term. The term $\frac{k}{a^2}$ corresponds to the spatial curvature on the brane. The ``$\pm$'' signs in Eq. (\ref{friedman0}) correspond to two different branches of the braneworld solutions. Here the lower sign ``$-$'' corresponds to the normal branch and the upper sign ``$+$'' corresponds to the self-accelerating branch.

In what follows, we will investigate the cosmological evolution of the model in a flat normal branch embedded in a flat bulk spacetime with vanishing bulk cosmological constant and without dark radiation (i.e. $k=0$, $\Lambda=0$ and $C=0$). We also consider the case where the brane tension $\sigma$ is taken to be positive. The negative brane tension usually demands the negative kinetic energy term leading to the ghost instability \cite{Burgess:2002vu,Parameswaran:2007cb,Abdalla:2010sz,Chen:2005jp}. The cosmological equation (\ref{friedman0}) then reduces to
\begin{equation}\label{friedman1}
H^2=\frac{\rho+\sigma}{3m^2}+\frac{2}{l^2}\Bigg[1-\sqrt{1+l^2\Big(\frac{\rho+\sigma}{3m^2}\Big)} \Bigg],
\end{equation}
or, equivalently
\begin{equation}\label{friedman2}
H=\sqrt{\frac{\rho+\sigma}{3m^2}+\frac{1}{l^2}}-\frac{1}{l}.
\end{equation}
Note here that the difference from the Friedmann equation in GR with a cosmological constant comes from the terms with $l$, which slow down the relative expansion rate. This difference disappears in the limit where the bulk piece in the action \eqref{1} vanishes and consequently $l\rightarrow \infty$.
The conservation equation of matter energy density is given by
 \begin{eqnarray}
   \dot{\rho}+3H(\rho+p)&=&0,\label{r_m_dot0}
 \end{eqnarray}
 where $p$ denotes the matter pressure and the over dot denotes derivative with respect to $t$.
The above cosmological equations are complemented by the acceleration equation
\begin{align}\label{acc_eqn}
\dot{H}=-\left(\frac{lH}{1+lH}\right) \left(\frac{\rho+p}{2 m^2}\right).
\end{align}

In order to determine the energy density contribution of different components, we define the following relevant energy density parameters viz. the relative matter energy density and the relative energy densities due to the brane gravity effects ($\Omega_{\sigma}$ and $\Omega_{l}$), respectively:
 \begin{eqnarray}\label{parameter0}
   \Om &=&\frac{\rho}{3m^2H^2},\nonumber \\
 \Omega_{\sigma} &=&\frac{\sigma}{3m^2H^2},\nonumber\\
  \Omega_{l} &=&-\frac{2}{l\,H}.
    \end{eqnarray}
The different relative energy density parameters are defined in such a way that they are related as
\begin{align}
\Omega_{\rm mat} + \Omega_{\sigma}+\Omega_l=1,
\end{align}
through Eq.~\eqref{friedman1}.
It may be noted that there is a possibility that $\Omega_l$ and $\Omega_\sigma$ range outside the interval $[0,1]$. In particular, for attractive gravity in the bulk and on the brane ($l>0$), thus for an  expanding universe we have from Eq. \eqref{parameter0}, $\Omega_l<0$.
From the same relation, we may define the effective energy density and pressure of the universe as
   \begin{eqnarray}
 \rho_{\rm eff} &=&\rho+\sigma-\frac{6m^2 H^2}{l H}, \nonumber \\
 p_{\rm eff} &=& - 3m^2 H^2+ \left(\frac{lH}{1+lH}\right) (p+\rho).
    \end{eqnarray}
Then the overall effective equation of state (EoS) is given by
    \begin{eqnarray} \label{weff_0}
 w_{\rm eff}&=&\frac{p_{\rm eff}}{\rho_{\rm eff}},\nonumber\\
 ~&=&-1+\Big(\frac{lH}{1+lH}\Big)\Big(\frac{\rho+p}{3 m^2\,H^2}\Big),
 \end{eqnarray}
 related to the deceleration parameter $q$ as
   \begin{eqnarray} \label{dec_0}
 q&=&-1-\frac{\dot H}{H^2}=\frac{1+3 w_{\rm eff}}{2}.
  \end{eqnarray}
For accelerated universe, we have the condition $q<0$ or $w_{\rm eff}<-\frac{1}{3}$. For an expanding universe,  $l>0$, and normal matter with $\rho\geq0$ and $p\geq0$, the effective barotropic index can not be phantom-like, $w_{\rm eff} \geq -1$, and superacceleration is not possible, $\dot{H}<0$. The limit $w_{\rm eff} \rightarrow -1$ is reached for dust matter ($p=0$) in the asymptotic future of the DE dominated era.

If we define the effective energy density and pressure of DE on the brane as follows \cite{Sahni:2006pa}
\begin{align}
\rho_{\rm DE}=\frac{3H^2}{m^2}(1-\Om), \qquad
p_{\rm DE}=\frac{2H^2}{m^2}(q-\frac{1}{2}),
\end{align}
then the effective EoS of DE $w_{\rm DE}=p_{\rm DE}/\rho_{\rm DE}$ is defined as
\begin{align}\label{w_de}
w_{\rm DE}=\frac{2q-1}{3(1-\Omega_{\rm mat})}= \frac{ w_{\rm eff} }{1-\Om} = \frac{ w_{\rm eff} }{\Omega_\sigma+\Omega_l} \,.
\end{align}
This result explains how in the late times when $\Om<1$ and $w_{\rm eff}\rightarrow -1 $, the brane construction manifests itself as phantom DE, $w_{\rm DE}<-1$. Of course, describing the brane effects in terms of a dark fluid is just a mathematical device. For example, since the relative densities of the brane effects scale differently in $H$, see Eq.~\eqref{parameter0}, and $\Omega_l<0$, the dark energy barotropic index $w_{\rm DE}$ will undergo a singularity at $H=\frac{\sigma l}{6 m^2}$ \cite{Viznyuk:2018eiz}. However, such abrupt change in the behavior of the dark fluid does not cause an anomaly in  the expansion of the universe and the overall effective barotropic index evolves smoothly. In the same context one may also note that during the earlier stages of the cosmological evolution when $|\Omega_l|>|\Omega_\sigma|$ it happens that $\Omega_{\rm mat}>1$ \cite{Viznyuk:2018eiz}, meaning that in the first part of the matter domination era the dust matter appears to be relatively overdense, when compared to the $\Lambda \rm CDM$ evolution.

In the following two sections, we will analyze the cosmological behavior of the above braneworld model in the case where there is no scalar field and the case where  a scalar field is trapped on the brane.

 \section{Stability analysis and cosmological behavior of critical points without scalar field}\label{sec:stability_pts_no_scalar}
In this section, we will first consider the case where there is no scalar field and the background fluid only consists of barotropic matter with energy density $\rho_{\rm mat}$ and pressure $p=w\rho_{\rm mat}$ present on the brane (i.e. $\rho=\rho_{\rm mat}$). Then the conservation equation \eqref{r_m_dot0} becomes
 \begin{eqnarray}
   \drm+3H(1+w)\rhm&=&0,\label{r_m_dot1}
 \end{eqnarray}
 where $w$ is the EoS of matter ($-1\leq w \leq 1$). In order to study the qualitative behavior of the model, we convert the above discussed cosmological equations into an autonomous system of equations. Then we shall use the dynamical system tools to properly extract the cosmological behavior of the model by analyzing the asymptotic behavior of critical points of the obtained autonomous system  without  digging into an almost impossible task of finding the analytical solution. It may be noted that cosmologically, a critical point represents the era for which the universe spent sufficient amount of time.

Hence, we introduce the following set of dimensionless phase space variables to recast the above cosmological equations into an autonomous system of differential equations
 \begin{equation}\label{variable1}
x= \frac{1}{lH},\qquad y = \frac{\sqrt{\sigma}}{\sqrt{3}mH}.
 \end{equation}
It may be noted here that the choice of variables plays a crucial role in determining the behavior of the model as different variables capture different aspects of the dynamics. For instance, for the case of the RS model, $m=0$, and thus the variable $y$ is undefined. Hence, the analysis of the autonomous system here does not cover the RS model and different dimensionless variables have to be considered (see Refs. \cite{Campos:2001cn,Iakubovskyi:2004ze}).

Using these dimensionless variables \eqref{variable1}, the cosmological equations \eqref{friedman2}, \eqref{acc_eqn} and \eqref{r_m_dot1} can be converted to the following autonomous system:
\begin{eqnarray}
x'&=&\frac{3(1+w)x}{2(1+x)}(1+2x-y^2),\label{yp}\\
y'&=&\frac{3(1+w)y}{2(1+x)}(1+2x-y^2),\label{zp}
\end{eqnarray}
where prime denotes differentiation with respect to a dimensionless variable $N=\ln a$.
Here, we should remark that since the variables $x$ and $y$ differ from each other by a constant factor in \eqref{variable1}, strictly speaking, the dynamics of the system is one dimensional. This property can be traced back to the fact that the original dynamical variables $H$ and $\rho$ are related via the Hubble constraint, Eq. \eqref{friedman2}. However, presenting the cosmological equations as the two-dimensional system \eqref{yp}-\eqref{zp}, helps us to keep track of the relative energy densities in a simple way, and capture the evolution of a whole family of models on a single 2-dimensional plot.

In terms of the variables \eqref{variable1}, the relative matter energy density parameter, the relative energy density parameters due to the brane gravity, the overall effective EoS and the effective EoS of DE are respectively given by
 \begin{eqnarray}\label{parameter}
  \Om &=& 1+2x-y^2, \qquad \Omega_{\sigma} = y^2, \qquad \Omega_{l} = -2x, \nonumber \\
  w_{\rm eff}&=&\frac {w(1+2x-y^2)+x-{y}^{2}}{1+x}, \qquad w_{\rm DE}={\frac {w(1+2x-y^2)+x-{y}^{2}}{ \left( 1+x \right)  \left({y}^{2}-2\,x \right) }}.
    \end{eqnarray}
The physical requirement condition $ \rhm \geq 0$  constrains the variables \eqref{variable1} to satisfy the inequality
\begin{equation}\label{constraint}
1+2x \geq y^2.
\end{equation}
Thus, the two dimensional phase space of the system (\ref{yp})-(\ref{zp}) is given by
\begin{equation}\label{phase}
\Psi=\Big\{(x,y) \in \mathbb{R}^2 :  y^2 \le 2x+1, x \geq 0 \Big \}.
\end{equation}

In order to obtain the qualitative behavior of the model governed by the system of equations (\ref{yp})-(\ref{zp}),
 we need to extract the critical points of the system by equating the left-hand side of equations to zero.
 The system is then perturbed in a neighborhood of each critical point and the stability of a critical point is determined by
 the nature of eigenvalues of the corresponding perturbed (Jacobian) matrix. According to the definition of $y$ in \eqref{variable1},
  the points with $y>0$ correspond to a positive value of $H$ (expanding universe), while points with $y<0$ correspond to  the negative value of $H$
  (contracting universe). However, the system \eqref{yp}-\eqref{zp} is invariant under the transformation  $y \rightarrow -y$. Hence, from physical as well as a mathematical point of view it is sufficient to analyze the system only for positive values of $y$. In the following two subsections, we will analyze the behavior of the critical points of the system \eqref{yp}-\eqref{zp} in the finite and infinite regions of the phase space.

\subsubsection{Analysis of finite critical points}

\begin{figure}
    \centering
    \subfigure[]{%
        \includegraphics[width=6cm,height=6cm]{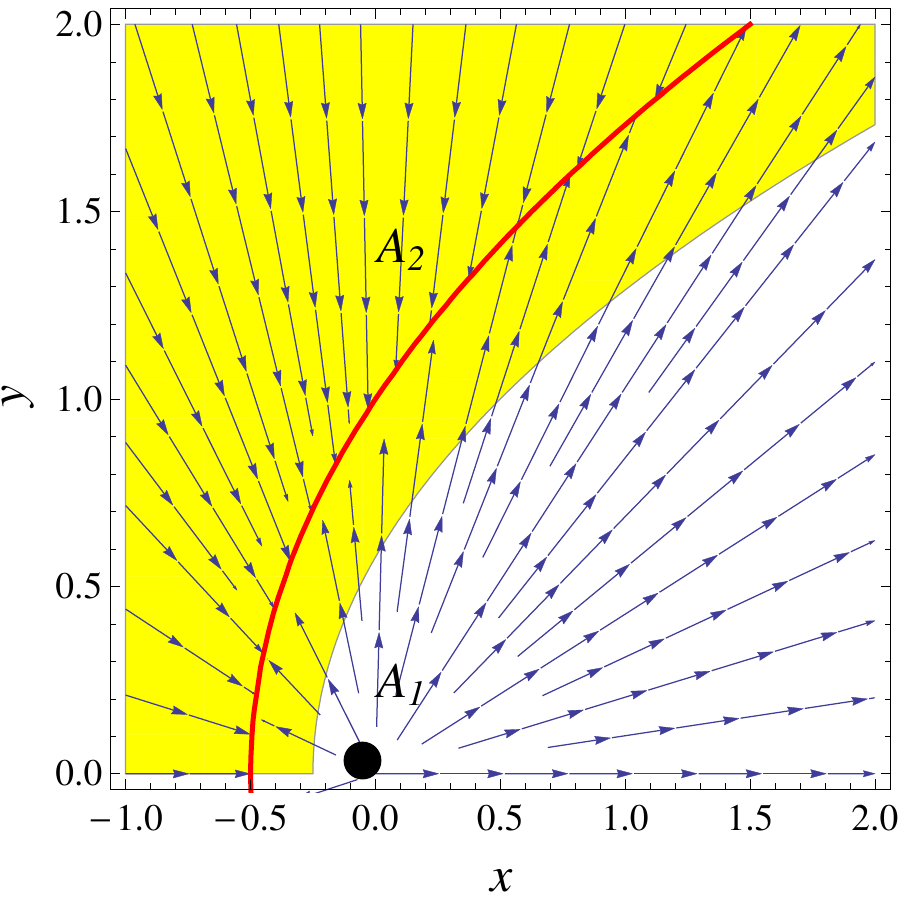}\label{fig:StreamPLot_C4}}
    \qquad
    \subfigure[]{%
        \includegraphics[width=6cm,height=6cm]{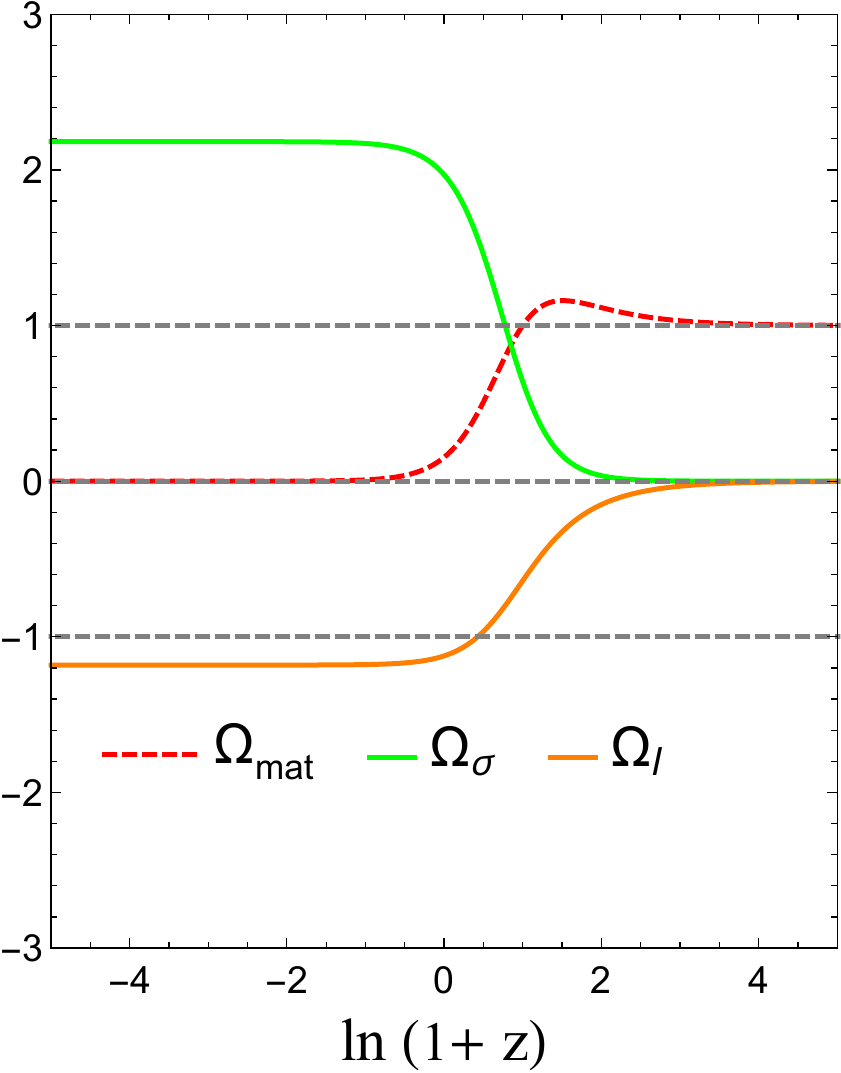}\label{fig:para_no_sca}}
        \qquad
    \subfigure[]{%
        \includegraphics[width=6cm,height=6cm]{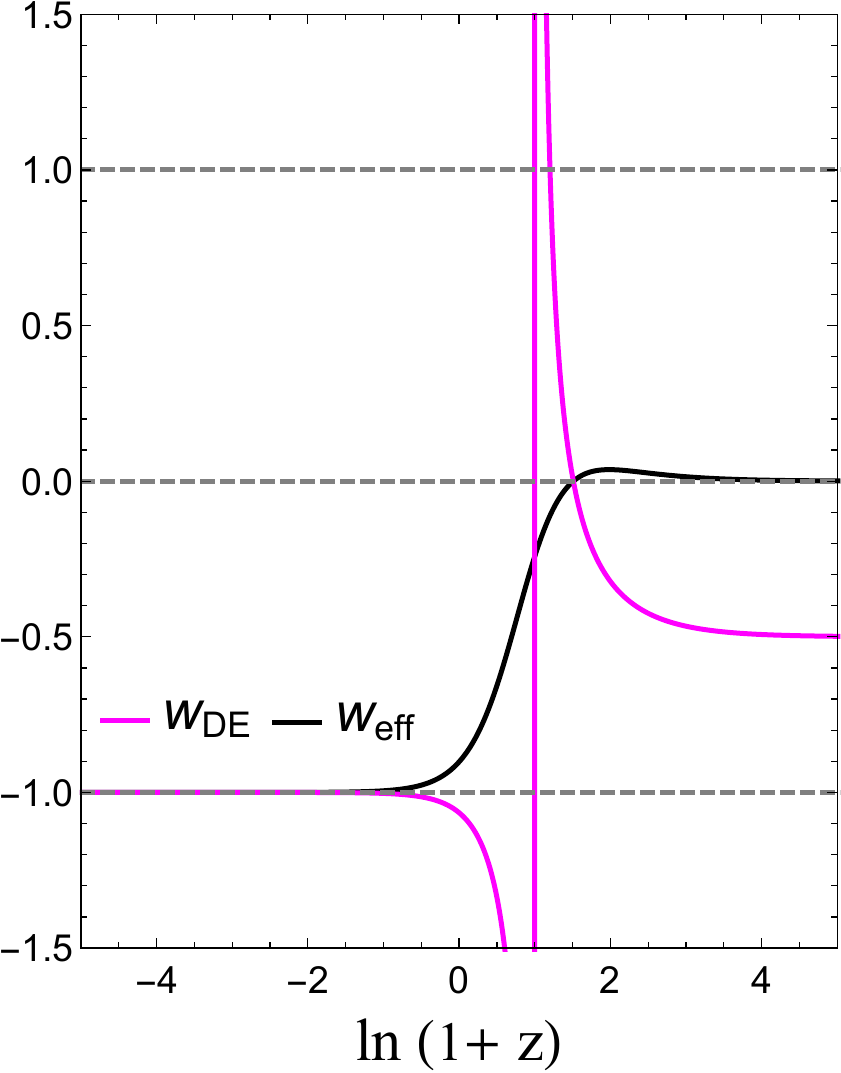}\label{fig:weff_no_sca}}
    \caption{(a) Phase space trajectories of the system \eqref{yp}-\eqref{zp} showing the stability nature of non-isolated set $A_2$.
     The yellow region represents the accelerated region and the red curve represents the curve $A_2$. (b) The time evolution of the relative matter energy density $\Omega_{\rm mat}$, the relative energy density due to brane corrections, $\Omega_\sigma$ and $\Omega_l$. (c) The time evolution of the effective EoS $w_{\rm eff}$ and the effective EoS of DE $w_{\rm DE}$. Here $w=0$. }
    \label{fig:c2_para_no_sca}
\end{figure}

For this case, there are only one critical point $A_1\,(0,0)$ and one set of critical points $A_2\,(x,\sqrt{2x+1})$, with $ x \geq0 $.
\begin{itemize}
\item The critical point $A_1$ corresponds to a matter dominated universe ($\Omega_{\rm mat}=1$, $w_{\rm eff}=w$), it is an unstable node with eigenvalues $\frac{3}{2} (1+w), \, \frac{3}{2} (1+w)$.

\item The set of critical points $A_2$ corresponds to an accelerated universe with the energy density dominated by the brane contribution  ($w_{\rm eff}=-1$, $\Omega_\sigma=2x+1$, $\Omega_l=-2x$, $\Omega_{\rm mat}=0$). The one-dimensional set has only one zero eigenvalue and this type of set of critical points is known as a normally hyperbolic set (in general, it is an $n$-dimensional set of nonisolated critical points containing $n$ zero eigenvalues). The stability of this set depends on the signature of the nonzero eigenvalues \cite{coley:2003mj}. For this set, since the only nonzero eigenvalue is $-3(1+w)$, hence it is a late time attractor. This can also be seen numerically by plotting the phase trajectories of the system as shown in Fig. \ref{fig:StreamPLot_C4}.

\end{itemize}

The dynamics of the system is depicted on Fig.~\ref{fig:c2_para_no_sca}. The phase portrait \ref{fig:StreamPLot_C4} illustrates the properties of the fixed point  $A_1$ and fixed set $A_2$. {\bf Trajectories in the outer left region $y^2>2x+1$ correspond to $\rho<0$ and those on the negative side of the $x$-axis correspond to repulsive gravity ($l<0$). Therefore, they are not physically relevant but are shown for the sake of completeness (see Eq. \eqref{phase})}.  The Figs.~\ref{fig:para_no_sca} and \ref{fig:weff_no_sca} present a numerical plot of the evolution of one particular solution with dust matter ($w=0$), showing the relevant energy density parameters and the EoS parameters against the redshift  $z=-1+\frac{a_0}{a}$, where $a_0$ represents the present value of the scale factor (taken to be equal to $1$). Recall that redshift  $z=0$ represents the present time of the universe, while $z=-1$ represents its infinite future.

%Note that the lower half with $y<0$ belongs to contracting universe, it has not been shown intentionally  due to the symmetry of the system as explained before.

From the above, we see that the viable dynamics of this model is very simple. The Universe starts evolving from the vicinity of a decelerated matter dominated era (point $A_1$) and reaches an accelerated era dominated by the brane gravity effects (set $A_2$). Fig.~\ref{fig:weff_no_sca} shows how at some moment during the evolution of the universe the effective EoS of DE ($w_{\rm DE}$) diverges, but this does not cause any disturbance since the overall effective EoS  ($w_{\rm eff}$) remains finite and smooth. As explained at Eqs.\ \eqref{w_de} and \eqref{parameter}, it is possible to have $w_{\rm DE}<-1$, but $w_{\rm eff} \nless -1$. Exactly this is happening, in the late stages, the effective DE is effectively phantom as it converges to a cosmological constant regime $w_{\rm DE}\rightarrow -1$ from below. In this late epoch, DE dominates over matter density. In the phase space, the points on the set $A_2$ correspond to the condition $\Omega_{ \sigma}+ \Omega_{l} =1$, realized by different values of the model parameters $\sigma$, $l$, and characterized by the Hubble expansion rate
\begin{equation}
H_{A_2} = - \frac{m \pm \sqrt{m^2 + \tfrac{1}{3}l^2 \sigma}}{lm} \,.
\end{equation}
Any particular model with given $\sigma$, $l$ is described by a single trajectory in the phase space, but the same trajectory can arise from different suitable combinations of the model parameters $\sigma$, $l$. By studying only the expansion history an observer can not uniquely determine the values of $\sigma$ and $l$, but sees only their combined effect.

Further, Figs.~\ref{fig:para_no_sca} and \ref{fig:weff_no_sca} also  illustrate an interesting feature how for $l>0$ the matter domination era begins with relative matter overdensity,  $\Omega_{\rm mat}>1$ , balanced by the effective contribution of the brane gravity,  $\Omega_{\sigma}+\Omega_l<0$. The relative matter overdensity grows to a maximum value and then starts to drop, passing through $\Omega_{\rm mat}=1$ at the moment when the brane effects cancel each other, $\Omega_{\sigma}=-\Omega_l$, and $w_{\rm DE}$ diverges  \cite{Viznyuk:2018eiz}.

We note that the phase space \eqref{phase} corresponding to the dynamical system (\ref{yp})-(\ref{zp}) is non-compact. Hence,  there could be some critical points with interesting cosmological features in the infinite region of the phase space which are non-trivial from the global analysis perspective. In order to extract the cosmological behavior of those points, in what follows we shall analyze the behavior of the system near infinity too.

\subsubsection{Analysis of critical points near infinity}\label{sec:poincare}
For the analysis of critical points near infinity for the system (\ref{yp})-(\ref{zp}), we have to extend our analysis using the Poincar\'e central projection method \cite{slynch}. The main idea of this method is to identify points near infinity of $\mathbb{R}^n$ with points on the surface of the sphere $S^{n-1}$ known as the Poincar\'e sphere. The behavior of the system on the Poincar\'e sphere gives the complete picture of the dynamical system in $\mathbb{R}^n$. This method has been used extensively in the analysis of the global phase space of several cosmological models (see for e.g. \cite{Leon:2014rra, Giacomini:2017yuk,Cruz:2017ecg}).

For this we introduce the following Poincar\'e coordinates $x_r$, $y_r$ defined as:
\begin{equation}\label{23}
x_r=\frac{x}{\sqrt{1+x^2+y^2}}, \qquad y_r=\frac{y}{\sqrt{1+x^2+y^2}}.
\end{equation}
Employing these new coordinates, the dynamical system (\ref{yp})-(\ref{zp}) transforms to
\begin{eqnarray}
x'_r&=&-\frac {3 (1+w) x_r}{2 \sqrt {-{R_r}^{2}+1} \left( \sqrt {-{R_r}^{2}+1}+x_r \right) } \Big( 2\,\sqrt {-{R_r}^{2}+1}(x_r^3+x_r y_r^2-x_r)-{x_r}^{4}-3\,{x_r}^{2}{y_r}^{2}-2\,{y_r}^{4} \nonumber \\&&~{} +2\,{x_r}^{2}+3\,{y_r}^{2}-1 \Big), \label{y_P} \\
y'_r&=&-\frac {3 (1+w)\,y_r}{2\,\sqrt {-{R_r}^{2}+1} \left( \sqrt {-{R_r}^{2}+1}+x_r \right) } \Big(2\,\sqrt {-{R_r}^{2}+1} \left({x_r}^{3}+x_r{y_r}^{2}-x_r\right)-{x_r}^{4}-3\,{x_r}^{2}{y_r}^{2}-2\,{y_r}^{4} \nonumber \\&&~{} +2\,{x_r}^{2}+3\,{y_r}^{2}-1 \Big), \label{z_P}
\end{eqnarray}
where $R_r=\sqrt{{x_{r}}^{2}+{y_{r}}^{2}}$. It can be seen that the system \eqref{y_P}-\eqref{z_P} is invariant under the transformation $y_r \rightarrow -y_r$. Further, using the physical requirement condition $\rho_{\rm mat}\geq 0$, the constraint (\ref{constraint}) can be written in terms of Poincar\'e variables as
\begin{equation}\label{exist_poincare}
 {\frac {\sqrt {1-R_r^2}+2
x_{r}}{\sqrt {1-R_r^2}}} \geq {
\frac {{y_{r}}^{2}}{1-R_r^2}}.
\end{equation}
 Hence, the phase space of the system \eqref{y_P}-\eqref{z_P} is given by
\begin{align}
\Psi_r=\left\lbrace (x_r,y_r) \in\,\, [0,1] \times [-1,1]\,\, :  \left(x_r+\sqrt{1-R_r^2}\right)^2\geq R_r^2, R_r^2\leq 1\right\rbrace.
\end{align}

\begin{table}[t]
\centering
\begin{tabular}{|c|c|c|c|c|c|c|c|c|c|c|c|}
        \hline
        Point~~~~&$x_r$~~~~&$ y_r$~~~~&~~~~~$\Omega_{\rm mat} ~$&~~~~$\Omega_{\sigma}~$&~~~~$\Omega_{l}$&~~~~$w_{\rm eff}$&~~~$w_{\rm DE}$~~~&~~~$\lambda_1$~~~&~~~$\lambda_2$\\\hline
        \hline
        $A^\infty$&$1$~&$0$~&$+\infty$&$0$&$-\infty$&$1$&$0$&$-6$&$0$\\\hline
\end{tabular}
\vspace{0.5cm}
\caption{Critical set of the system (\ref{y_P})-(\ref{z_P}) and values of the relevant cosmological parameters for general potential along with the corresponding eigenvalues.} \label{Table_exp_inf}
\end{table}

\begin{figure}
    \centering
            \includegraphics[width=6cm,height=6cm]{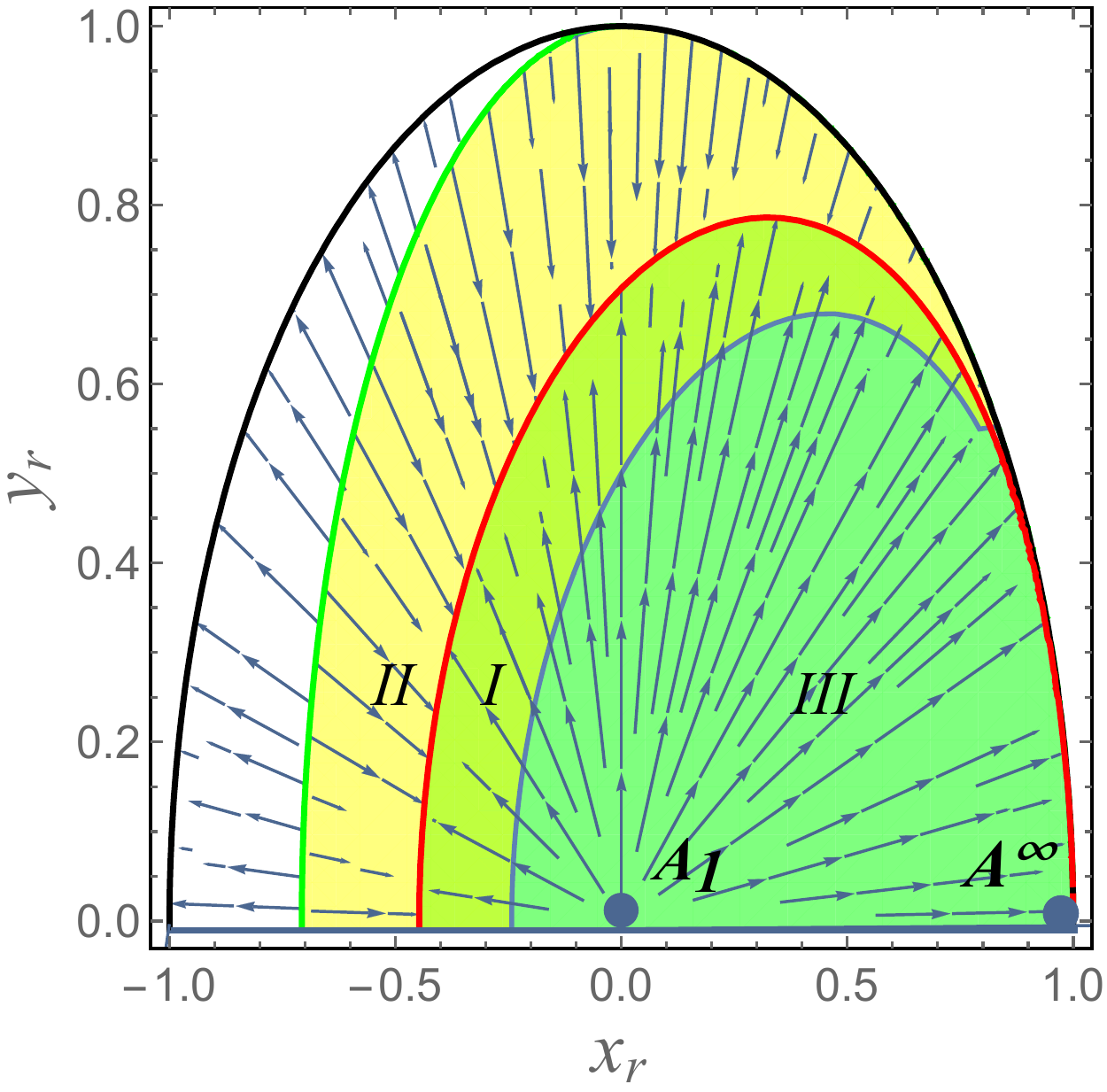}
            \caption{Phase space trajectories of the system \eqref{y_P}-\eqref{z_P}. The areas I and II represent the region of
acceleration whereas the areas I and III represent the physical phase space of the system. The red colored
curve represents the set $A_2$. The green colored curve represents the curve where the system \eqref{y_P}-\eqref{z_P} becomes singular. Here $w=0$.
% \wk{Fig. has been redrawn due to the symmetry $y_r \rightarrow -y_r$.}
}  \label{fig:Poincare_stream_plot_y_z}
\end{figure}

\begin{figure}
    \centering
        \includegraphics[width=6cm,height=6cm]{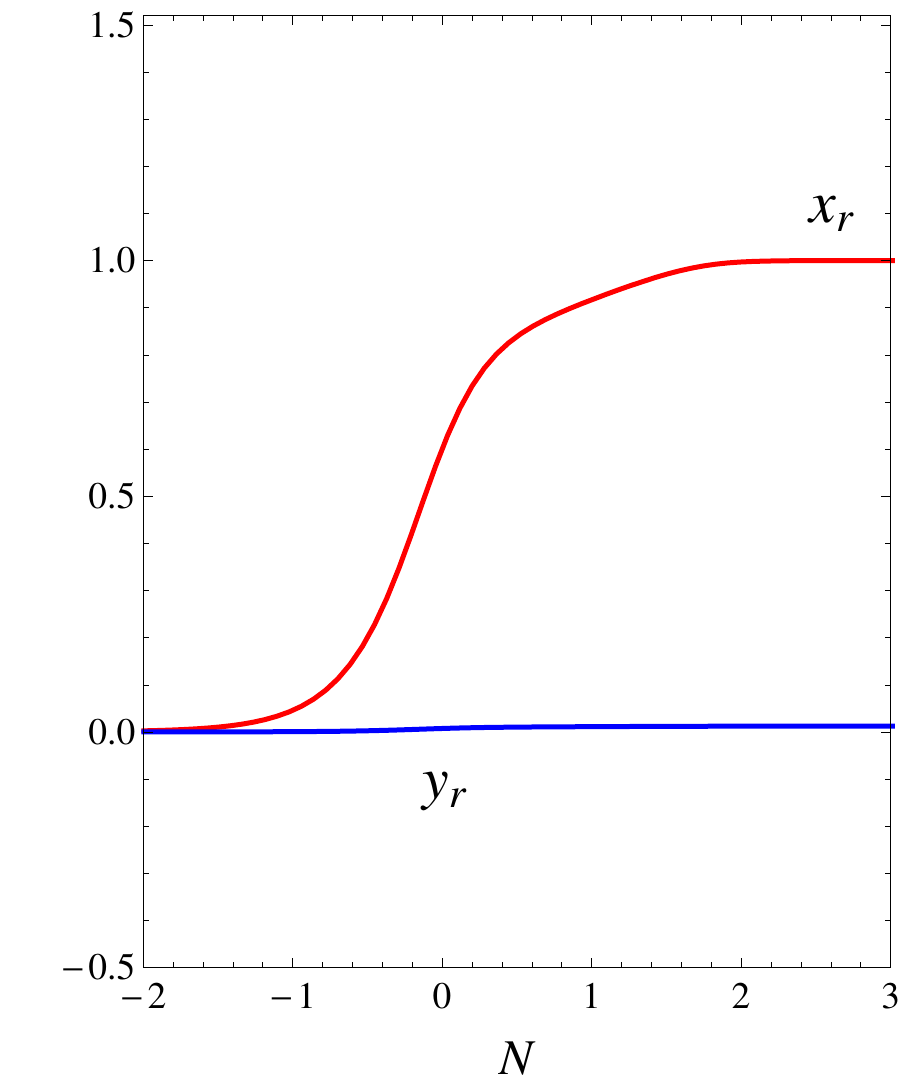}\label{fig:pert_A_inf_exp}
        \caption{The evolution of the Poincar\'e coordinates of the solution of the system \eqref{y_P}-\eqref{z_P}, depicting the universe with vanishing brane tension $\sigma$, which evolves from the point $A_1$ to the point $A^\infty$.}\label{fig:Poincare_pert_plot_exp}
\end{figure}

It can be seen that the system  \eqref{y_P}-\eqref{z_P} possess a boundary of the upper part of the circle $x_r^2+y_r^2=1$ as a set  of critical points at infinity. However, only a point $A^{\infty} (1,0)$ is physical i.e.\ belongs to the physical phase space. The values of the cosmological parameters at this point, as well as the corresponding eigenvalues, are given in Table \ref{Table_exp_inf}. This point exhibits a stiff matter dominated solution $w_{\rm eff}=1$ and $\Omega_{\rm mat} \rightarrow +\infty$. Hence, this point is not viable for a late universe.  By linearizing the system near this point, it can be seen that this point is nonhyperbolic with a 1-dimensional stable manifold.  The  phase space trajectories of the system \eqref{y_P}\eqref{z_P} are shown in Fig. \ref{fig:Poincare_stream_plot_y_z}, it is found that trajectories in vicinity of point $A^\infty$ approach the boundary of the semicircle of circle $x_r^2+y_r^2=1$ which contains point $A^\infty$. The solution which evolves from a matter dominated point $A_1$ towards a point  $A^\infty$ is unique and corresponds to the absence of brane tension ($y_r=0$). This behavior can also be illustrated by plotting the projections of the system \eqref{y_P}-\eqref{z_P} on $x_r$, $y_r$ axes as shown in Fig. \ref{fig:Poincare_pert_plot_exp}. Since the point is nonhyperbolic, the stability of such a point is usually analyzed using the center manifold theory. However, since this point is not of cosmological interest we will not study it further and present it only for the sake of completeness.

Thus, from the above analysis, we can summarize that the viable cosmological behavior can be described only by the finite critical points.

\section{Stability analysis and cosmological behavior of critical points with scalar field}  \label{sec:stability_pts}
In this section, we will consider the case where the background fluid consists of barotropic matter along with the scalar field $\phi$ trapped on the brane. Scalar fields are very simple to deal with and can play a useful role in the cosmological dynamics capable to mimic the cosmological constant at late times. They are used to describe inflation and explain the possibility of the graceful exit from the inflationary era. The energy density and pressure of the scalar field are respectively given by
  \begin{equation}\label{phi_dens_pres}
 \rho_\phi=\frac{\dot \phi^2}{2}+V(\phi), \hspace{0.4cm} p_\phi=\frac{\dot \phi^2}{2}-V(\phi).
 \end{equation}
 Here, $V(\phi)$ is the self-interacting potential for the scalar field $\phi$. The total energy density is then given by $\rho= \rhm+\rho_\phi$ and hence the cosmological equations \eqref{friedman2} and \eqref{acc_eqn} can be written as
\begin{equation}\label{friedman2_sca}
H=\sqrt{\frac{\rho_{\rm mat}+\rho_{\phi}+\sigma}{3m^2} + \frac{1}{\l^2}}-\frac{1}{l}.
\end{equation}
and
 \begin{align}\label{acc_eqn_sca}
\dot{H}=-\left(\frac{lH}{1+lH}\right) \left(\frac{(1+w)\rho_{\rm mat}+\rho_\phi+p_\phi}{2 m^2}\right).
\end{align}
We also assume that the energy densities for scalar field and matter are separately conserved given by
\begin{eqnarray}
  \dot \rho_\phi+3H\rho_\phi(1+w_\phi)&=&0,\label{r_phi_dot}\\
  \drm+3H(1+w)\rhm&=&0,\label{r_m_dot}
 \end{eqnarray}
 where $w_\phi =\frac{p_\phi}{\rho_\phi}$ is the EoS for scalar field $\phi$. Combining \eqref{phi_dens_pres} and \eqref{r_phi_dot}, the evolution equation of the scalar field is then given by
 \begin{equation}\label{KG_phi}
  \ddot \phi=-3H\dot \phi-\frac{dV}{d\phi}.
  \end{equation}

We now introduce the following set of dimensionless phase space variables
 \begin{equation}\label{variable2}
x= \frac{1}{lH},\qquad y = \frac{\sqrt{\sigma}}{\sqrt{3}mH},\qquad u = \frac{\dot{\phi}}{\sqrt{6}mH},\qquad v = \frac{\sqrt{V}}{\sqrt{3}mH},\qquad s=-\frac{m}{V}\frac{dV}{d \phi},
 \end{equation}
to recast the above cosmological equations into an autonomous system of differential equations:
\begin{eqnarray}
x'&=&\frac{3xu^2}{1+x}+\frac{3x(1+w)}{2(1+x)}(1+2x-y^2-u^2-v^2),\label{y_p}\\
y'&=&\frac{3yu^2}{1+x}+\frac{3y(1+w)}{2(1+x)}(1+2x-y^2-u^2-v^2),\label{z_p}\\
u'&=&-3u+\sqrt{\frac{3}{2}}s v^2+u\Bigg[\frac{3u^2}{1+x}+\frac{3(1+w)}{2(1+x)}(1+2x-y^2-u^2-v^2)\Bigg],\label{u_p}\\
v'&=&-\sqrt{\frac{3}{2}}s uv+v\Bigg[\frac{3u^2}{1+x}+\frac{3(1+w)}{2(1+x)}(1+2x-y^2-u^2-v^2)\Bigg],\label{v_p}\\
s'&=&-\sqrt{6} u f(s), \label{s_p}
\end{eqnarray}
where we have defined
\begin{align}\label{Gamma}
f(s)=s^2 \left(\Gamma(s)-1\right)~~~ \text{with}~~~~ \Gamma=V\frac{d^2V}{d \phi^2}\Big/ \Big(\frac{dV}{d \phi}\Big)^2.
\end{align}

\begin{table}%[!ht]
\centering
\begin{tabular}{|c|c|c|c|}
  \hline
  $V(\phi)$ & ~~~~$f(s)$~~~~ & ~~~~$s_*$ ~~~~& References \\
  \hline
&&&\\

  $V_0\sinh^{-\eta}(\mu\phi)$&&&  \\
  & \raisebox{1ex} {$s^2/\eta-\eta \mu^2$}&  \raisebox{1ex} {$ \pm \eta\mu$}&  \raisebox{1ex} {\cite{Sahni:1999gb}} \\[-3ex]
 \raisebox{2.5ex} {$V_0\cosh^{-\eta}(\mu\phi)$}&~~&~~&\\[1.5ex]

     $\frac{V_0}{\phi^n}$ & $\frac{s^2}{n}$ & $0$ & \cite{Zlatev:1998tr} \\[2.5ex]%&No\\

   $\frac{V_0}{(\eta+e^{-\alpha\phi})^\beta}$ & $\frac{s^2}{\beta}+s\, \alpha$ & $0$ or $-\alpha\beta$ & \cite{Zhou:2007xp} \\[2ex]%&No

   $V_1 e^{\alpha\phi}+V_2 e^{\beta\phi}$ & $-(\alpha+s)(\beta+s)$ & $-\alpha$ or $-\beta$ & \cite{Jarv:2004uk} \\[2ex]%&No\\

   \hline
\end{tabular}
%\end{adjustbox}
\caption{The function $f(s)$ and $s_*$ for some common quintessence potentials.}
\label{tab:potential}
\end{table}

Different potentials will give different expression of the function $\Gamma$. For instance, the exponential potential is a simple and widely studied example with $\Gamma=1$ for any $s$.
In order to close the system \eqref{y_p}-\eqref{s_p}, in this work we consider only the potentials where $\Gamma$ can be written as a function of $s$ explicitly.
%In what follows, we shall confine our analysis to a category of potentials where $\Gamma$ can always be written as a function of $s$.
Since both $\Gamma$ and $s$ are functions of $\phi$, so in principle one can possibly relate them, i.e.\ if $s(\phi)$ is invertible we obtain $\phi(s)$ and hence we can write $\Gamma$ as function of $s$.
This assumption is valid for a wide class of scalar field potentials considered in the context of cosmology (see e.g.~\cite{Tamanini:2014nvd,Bahamonde:2017ize}).
In Table~\ref{tab:potential}, we present some important quintessence potentials along with expressions of the function $f(s)$ and values of $s_*$ (here $s_*$ is the solution of equation $f(s)=0$).

In terms of the variables (\ref{variable2}) the relative scalar field energy density, the relative matter energy density, and the relative energy densities due to the brane gravity effect ($\Omega_{\sigma}$ and $\Omega_{l}$), as well as the overall effective EoS and the effective EoS of DE are respectively given by
 \begin{eqnarray}\label{parameter2}
 \Omega_\phi &=&\frac{\rho_\phi}{3m^2H^2}=u^2+v^2,\nonumber \\
  \Om &=&1+2x-u^2-v^2-y^2,\nonumber \\
 \Omega_{\sigma} &=&y^2,\nonumber\\
  \Omega_{l} &=&-2x, \nonumber\\
  w_{\rm eff}&=&\frac {w(1-u^2-v^2-y^2+2x)+{u}^{2}-{v}^{2}-{y}^{2}+x}{1+x}, \nonumber\\
 w_{\rm DE}&=&{\frac {w(1-u^2-v^2-y^2+2x)+{u}^{2}-{v}^{2}-{y}^{2}+x}{ \left( 1+x \right)  \left( {u}^{2}
+{v}^{2}+{y}^{2}-2\,x \right) }}.
    \end{eqnarray}
Note that,  here scalar field, as well as brane gravity terms, contribute to DE. The different relative energy density parameters are connected by the relation
\begin{align}
\label{sum of Omegas with scalar field}
\Omega_{\sigma}+\Omega_l+\Omega_\phi+\Omega_{\rm mat}=1.
\end{align}

The physical requirement condition $ \rhm \geq 0$  constrains the variables \eqref{variable2} to satisfy the inequality
\begin{equation}\label{constraint2}
1+2x \geq (y^2+u^2+v^2),
\end{equation}
and hence the five dimensional phase space of the system (\ref{y_p})-(\ref{s_p}) is given by
\begin{equation}\label{phase2}
\Psi=\Big\{(x,y,u,v) \in \mathbb{R}^4 :  y^2+u^2+v^2 \le 2x+1, x \geq 0 \Big \}\times \left\lbrace s \in \mathbb{R}\right\rbrace.
\end{equation}
Similar to the case in the Sec.\ \ref{sec:stability_pts_no_scalar}, from the definition of $y$ and $v$ in \eqref{variable2}, the points with $y>0$ and $v>0$ correspond to a positive value of $H$ (expanding universe), while points with $y<0$ and $v<0$ correspond to a negative value of $H$ (contracting universe). However, the system \eqref{y_p}-\eqref{s_p} is invariant under the transformation  $y \rightarrow -y$ and $v \rightarrow -v$. Hence, we will focus only upon the positive values of $y$ and $v$ (expanding universe). It is also noted that the analysis of this model coincides with the analysis of DGP case \cite{Quiros:2008hv} by setting the variable $y=0$.  The analysis here also coincides with the analysis of the canonical scalar field in GR by considering the variable $x=y=0$ \cite{Copeland:1997et}.

\begin{table}
\centering
\begin{tabular}{|c|c|c|c|c|c|c|c|c|c|c|c|c|}
        \hline
        Point~~~~&$x$~~~~&$ y$~~~~&~~~$u$~~~&~~~$v$&$s$~~~~&~~~~${\Omega}_{\phi}~$&~~~~$\Om~$&~~~~$\Omega_{\sigma}~$&~~~~$\Omega_{l}$&~~~~$w_{\rm eff}$&~~~$w_{\rm DE}$\\\hline
        \hline
        $B_1$&$0$~&$0$~&$0$~&$0$~&$s$~&$0$~&$1$&$0$&$0$&$w$&undefined\\\hline
        $B_{2\pm}$&$0$~&$0$~&$\pm 1$~&$0$~&$s_*$&~$1$~&$0$~&$0$&$0$&$1$&$1$ \\\hline
        $B_3$&$0$~&$0$~&$\frac{s_*}{\sqrt{6}}$~&$\sqrt{1-\frac{s_*^2}{6}}$&$s_*$~&$1$~&$0$~&$0$&$0$&$\frac{s_*^2}{3}-1$&$\frac{s_*^2}{3}-1$\\\hline
        $B_4$&$x_c$~&$\sqrt{2x_c+1}$~&$0$~&$0$&$s_c$~&$0$~&$0$~&$2x_c+1$&$-2x_c$&$-1$&$-1$~\\\hline
        $B_5$&$0$~&$0$~&$\frac{1}{2}{\frac {\sqrt {6} \left(w+1 \right) }{s_*}}$~&$\frac{1}{2}{\frac {\sqrt {6(1-{w}^{2})}}{s_*}}$~&$s_*$~&$\frac{3(w+1)}{s_*^2}$~&$1-\frac{3(w+1)}{s_*^2}$~&$0$&$0$&$w$&$\frac{w s_*^2}{3(w+1)}$~\\\hline

\end{tabular}
\vspace{0.5cm}
\caption{The critical points of the system (\ref{y_p})-(\ref{s_p}) and values of the relevant cosmological parameters. {\bf Here $x_c$ and $s_c$ denote an arbitrary value of $x$ and $s$ respectively, where $s_c \in \mathbb{R}$  and $x_c \geq 0$.}} \label{Table1}
\end{table}

\begin{table}
\centering
\begin{tabular}{|c|c|c|c|c|c|c|}
        \hline
        Point~~~&~~~$\lambda_1$~~~&~~~$\lambda_2$~~~&~~~$\lambda_3$~~~~&~~~~$\lambda_4$&$\lambda_5$~~~~\\\hline
        \hline
        $B_1$~&$\frac{3}{2}(w-1)$&$\frac{3}{2}(w+1)$&$\frac{3}{2}(w+1)$&$\frac{3}{2}(w+1)$&$0$\\\hline
        $B_{2\pm}$~&$3\mp \frac{\sqrt{6}}{2}s_*$~&$3(1-w)$&$3$&$3$&$\mp \sqrt{6}df(s_*)$ \\\hline
        $B_3$~&$s_*^2-3(1+w)$~&$\frac{s_*^2}{2}-3$&$\frac{s_*^2}{2}$&$\frac{s_*^2}{2}$&$-s_*df(s_*)$ \\\hline
        $B_4$&$-3(w+1)$~&$-3$&$0$&$0$&$0$\\\hline
        $B_5$~&$\delta_+$~&$\delta_-$&$\frac{3}{2}(w+1)$&$\frac{3}{2}(w+1)$&$-\frac{3(w+1)df(s_*)}{s_*}$\\\hline

            \end{tabular}
\vspace{0.5cm}
\caption{ Eigenvalues of critical points given in Table \ref{Table1}. Here, $\delta_\pm=\frac{3}{4}\Big[(w-1)\pm \frac{1}{s_*}\sqrt{(w-1)\left((7+9w)s_*^2-24(w+1)^2\right)}\Big]$} \label{Table2}
\end{table}

The system \eqref{y_p}-\eqref{s_p} contains five critical points $B_1$, $B_{2\pm}$, $B_3$, $B_5$ and one set of non-isolated critical points $B_4$ in the finite region of the phase space \eqref{phase2} (see  Tables \ref{Table1} and \ref{Table2}). Out of all six critical points, only the set of critical points $B_4$ corresponds to an effect from brane gravity corrections. The critical point $B_1$ and the set $B_4$ are independent of the choice of potential for their existence, however, the stability of $B_4$ depends on the choice of potential. The critical points $B_{2\pm}$, $B_3$ and $B_5$ depend on the choice of the  potential and the number of copies of these points runs as the number of solutions $s_*$ of $f(s)=0$ . Throughout the paper, we use the notation $df$ to denote the derivative of $f$ with respect to $s$. We now briefly discuss the stability and the cosmological behavior of the critical points listed in Table \ref{Table1}.

\begin{figure}
    \centering
            \includegraphics[width=6cm,height=5cm]{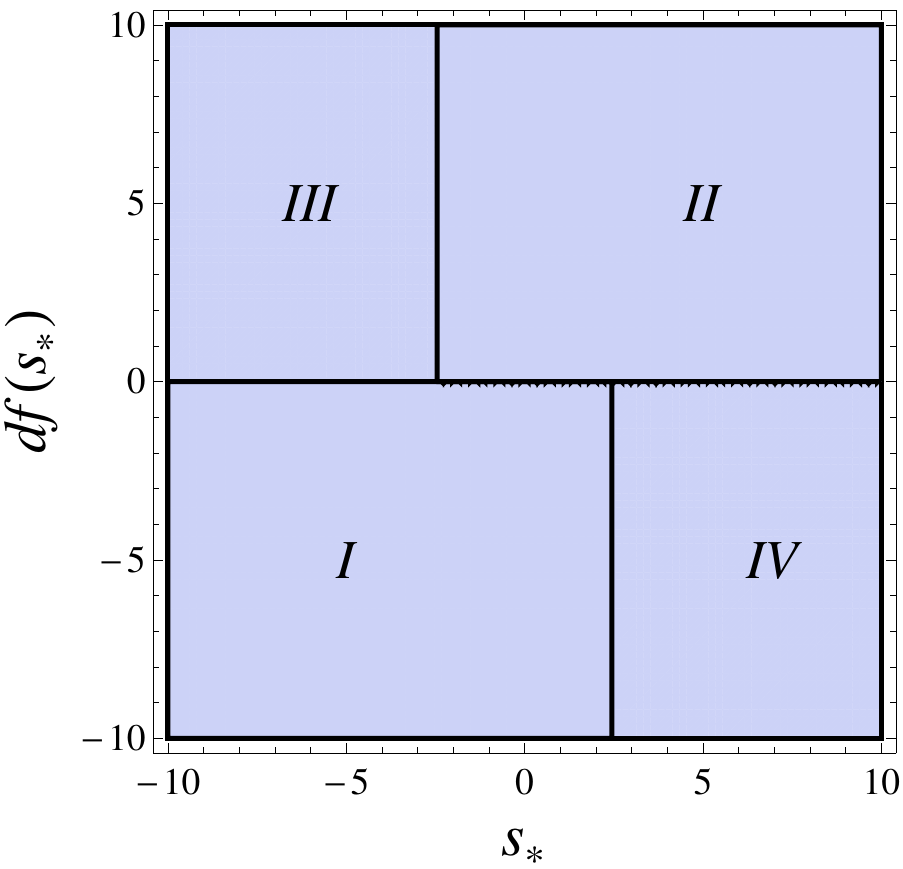}
        \caption{Regions of unstable node and saddle of point $B_{2\pm}$ on $(s_*,df(s_*))$ parameter space. Region $I$ represents the region of unstable node of $B_{2+}$ and regions $II$, $III$ and $IV$ represent the regions of saddle of $B_{2+}$. Region $II$ represents the region of unstable node of $B_{2-}$ and regions $I$, $III$ and $IV$ represent the regions of saddle of $B_{2-}$ .}
    \label{fig:c2pm_region_gen}
    \end{figure}

    \begin{figure}
    \centering
            \includegraphics[width=6cm,height=5cm]{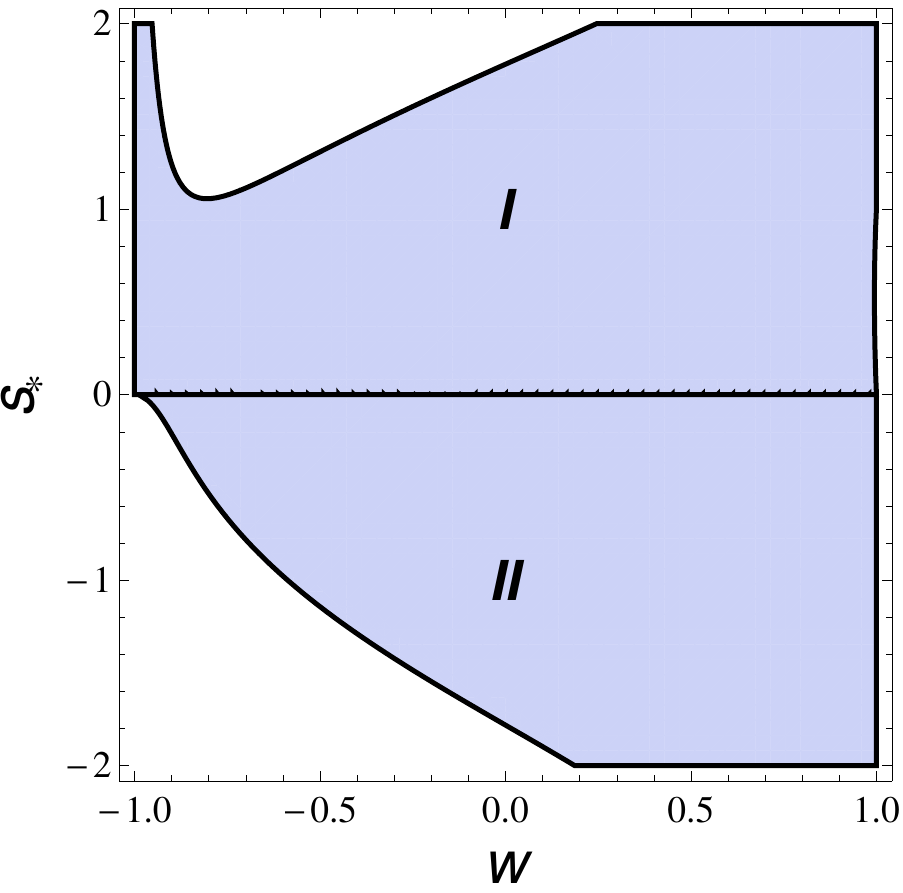}
        \caption{Regions where eigenvalues $\lambda_1$ and $\lambda_2$ of point $B_5$ are positive on $(w,s_*)$ parameter space. Region $I$ represents region where real part of $\lambda_1$ is positive and region $II$ represents region where real part of $\lambda_2$ is positive.}
    \label{fig:Region_Plot_B5}
    \end{figure}

\begin{itemize}

\item The critical point $B_1$ corresponds to an unaccelerated matter dominated universe ($\Om =1$, $w_{\rm eff}=w$) even though the effective DE is undetermined.

\item The critical points $B_{2\pm}$ correspond to stiff matter like solutions $(w_{\rm eff}=1)$ dominated by the kinetic term of the scalar field. This point does not correspond to an accelerated universe. Point $B_{2+}$ is a past time attractor if $s_*<\sqrt{6}$ and $df(s_*)<0$ otherwise it is a saddle, whereas point $B_{2-}$ is a past time attractor if $s_*>-\sqrt{6}$ and $df(s_*)>0$ otherwise it is a saddle. Fig. \ref{fig:c2pm_region_gen} depicts the region in $(s_*,df(s_*))$ parameter space for which these points behaves as saddle and unstable node.

\item The point $B_3$ corresponds to a scalar field dominated solution and exists for $s_*^2<6$. It corresponds to an accelerated universe for $s_*^2<2$. From the nature of eigenvalues, this point is always a saddle for $s_* \neq 0$ and any value of $w$ (since $\lambda_2<0$ and $\lambda_3$, $\lambda_4>0$) which is contrasting with its stability behavior in standard GR case. However, it is a nonhyperbolic point for specific potential with $s_*=0$, in which case one has to apply center manifold theory. For simplicity, we postpone the analysis to the case of a specific potential.

\item The brane gravity dominated solution described by the set $B_4$ corresponds to an accelerating universe. It is non-hyperbolic in nature, so linear stability theory fails. Hence, further investigation is required to determine the dynamical behavior of the center manifold near this set. The complete stability analysis near the center manifold of a set $B_4$ is presented in the appendix \ref{app:1} along with the detailed stability conditions. Although this set is independent of the choice of the potential for its existence,  however, it can either behave as a saddle or stable set depending on the values of $s_c$ i.e.\ on the choice of potential and on the barotropic fluid's equation of state $w$. Since the stability conditions of points on this set depending on the choice of potential (see Appendix \ref{app:1}), we shall further find the conditions of the specific potential parameters by analyzing for a specific choice of the potential. {\bf Note that this set of critical points also appears in the case without scalar field implying the generic behavior of the value of $x_c$, unlike $s_c$, which depends on the choice of scalar field potential.}

\item The critical point $B_5$ exists for $s_*^2<3$ and $s_* \neq 0$. It corresponds to a scaling solution where the DE density  scales as the background energy density. Usually, a scaling solution can alleviate the coincidence problem, however, this point corresponds to an unaccelerated solution while the Universe expands as if it is matter dominated ($w_{\rm eff}=w$). From Table \ref{Table2}, it can be seen that for any choice of $w$ and $s_*$, the real parts of the eigenvalues $\lambda_1$ and $\lambda_2$ cannot be both positive (see Fig. \ref{fig:Region_Plot_B5}) but the eigenvalues $\lambda_3$, $\lambda_4$ are both positive. Hence, the point $B_5$ is a saddle in nature and thus corresponds to an intermediate period of the Universe.
\end{itemize}

We have also checked that as in the case of absence of the scalar field, the system contains only one point near infinity with stiff matter domination. However, since this is not interesting from the late time perspective, we will not analyze its stability in detail. As the behavior of few finite critical points depends on the choice the potential, it is, therefore, better to analyze the dynamics of the system \eqref{y_p}-\eqref{s_p} for some specific choice of the potential.  In the remaining part of this section, we consider three choices of the potential as examples which belong to three distinct categories:
\begin{itemize}
\item The potential where $\Gamma(s)=1$ for all $s$ for e.g.\ the exponential potential.
\item The potential where $\Gamma(s)=1$ for some values of $s$ for e.g.\ the hyperbolic potential.
\item The potential where $\Gamma(s) \neq 1$ for all values of $s$ for e.g.\ the inverse power-law potential.
\end{itemize}

\subsection{Exponential potential $V=V_0 \exp ^{-\frac{\lambda \phi}{m}}$}\label{subsec:exp}

In this section, we consider the potential
\begin{align}
V=V_0 \exp ^{-\frac{\lambda \phi}{m}},
\end{align}
where $V_0$ and $\lambda$ are constants of suitable dimension. It is worth mentioning that this is the only potential belonging to the category of potentials where $\Gamma=1$. Apart from its mathematical simplicity, the exponential potential of the scalar field model can be motivated from fundamental theories such as String theory/ M-Theory. Exponential potential has been also considered to model the cosmological inflationary period \cite{Lucchin:1984yf}.

For this choice of the potential, the variable $s$ is fixed (i.e. $s=\lambda$) and hence the system (\ref{y_p})-(\ref{s_p}) reduces to four dimensions. Since the critical point $B_1$ is independent of the choice of the potential, hence the stability and cosmological behavior remain the same as in the general potential case. Stiff matter dominated critical points $B_{2\pm}$  are past time attractors if $\pm \lambda<\sqrt{6}$, otherwise they are saddles. The scalar field dominated critical point $B_3$ exists for $\lambda^2<6$ and corresponds to an accelerated universe for $\lambda^2<2$. However, for this potential, this point is always a saddle in nature, hence it cannot describe the late universe. As we have seen for the general potential case, the accelerated, brane gravity dominated set of critical points $B_4$ is  non-hyperbolic in nature. Since for this potential, the variable $s$ is constant, hence from the analysis in the appendix \ref{app:1}, we can conclude for $w=0$ case,this set always behaves as a late time attractor for any choice of $\lambda$. However, for $w\neq 0$, the points on this set behaves as stable points if $\frac{w}{x_c+1}>0$. This result is interesting as it shows how the matter content affects the brane gravity domination over late time universe. As we have seen in the general potential case, the scaling solution $B_5$ corresponds to an unaccelerated solution and it is saddle in nature for any values of $\lambda$ and $w$.

\begin{figure}
    \centering
    \subfigure[]{%
        \includegraphics[width=6cm,height=5cm]{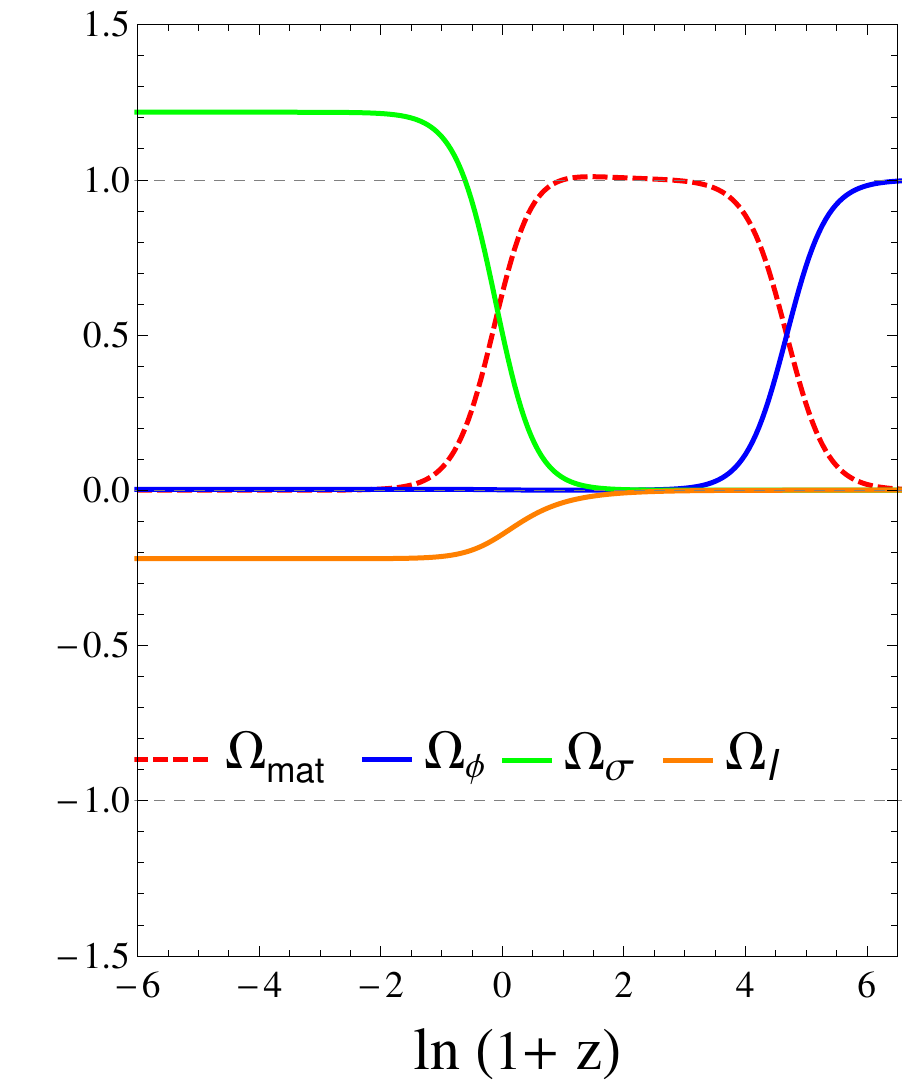}\label{fig:para_sca_1}}
    \qquad
    \subfigure[]{%
        \includegraphics[width=6cm,height=5cm]{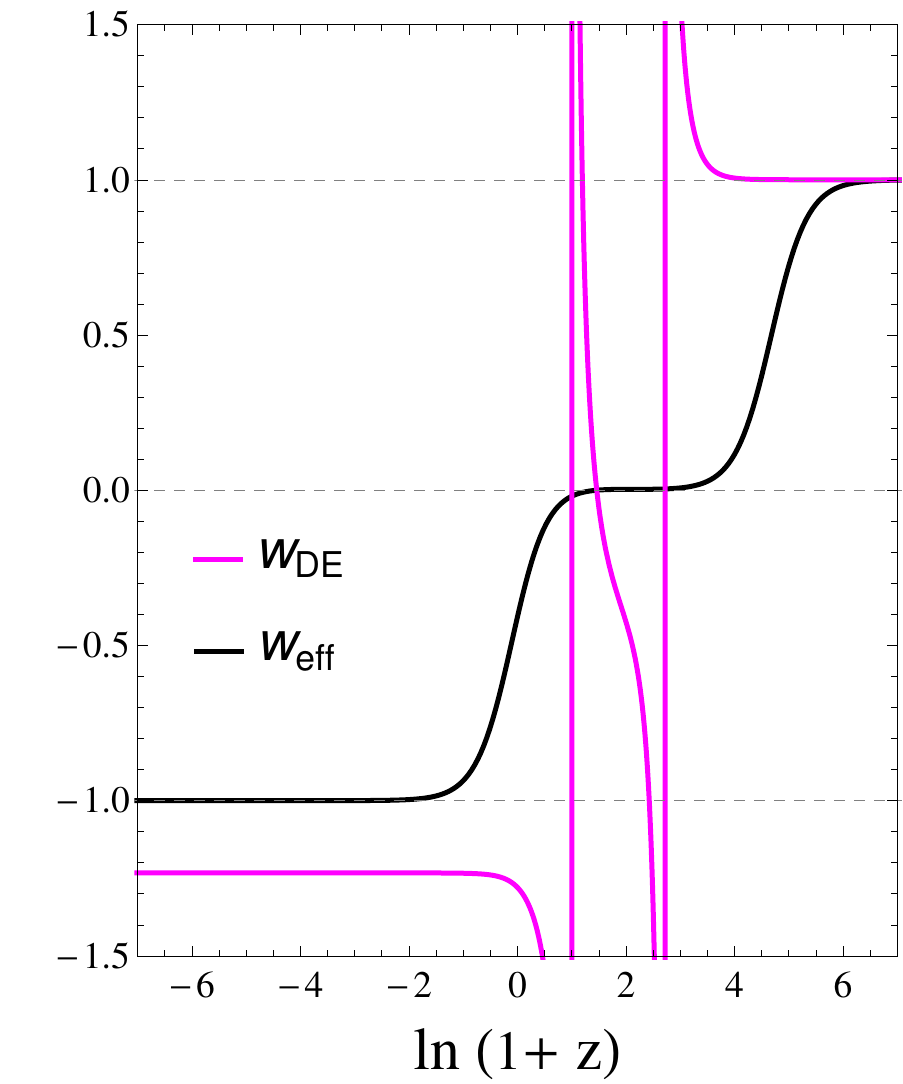}\label{fig:w_eff_sca}}
    \caption{(a) The time evolution of the relative matter energy density $\Omega_{\rm mat}$, the relative DE density $\Omega_\phi$ and the relative energy density due to brane corrections $\Omega_\sigma$ and $\Omega_l$. (b) The time evolution of the effective EoS $w_{\rm eff}$ and the effective EoS of DE $w_{\rm DE}$. In both panels, we take the potential $V=V_0 \exp ^{-\frac{\lambda \phi}{m}}$ with $\lambda=0.1$ and $w=0$.}
    \label{fig:para_w_eff_sca}
    \end{figure}

Thus depending on the choice of the model parameters, the universe evolves from a stiff matter dominated era $B_{2\pm}$ ($\Omega_\phi=1,\, w_{\rm eff}=1$) towards  an accelerating era dominated by the energy density due to the brane effects $B_4$ ($\Omega_\sigma+\Omega_l=1$, $w_{\rm eff}=-1$) either through a matter domination $B_1$ ($\Omega_{\rm mat}=1$, $w_{\rm eff}=w$) or unaccelerated scaling solution $B_5$ or DE dominated solution $B_3$ ($\Omega_\phi=1$, $w_{\rm eff}=-1$). Moreover, it can be seen that for the case $w=0$ without any fine-tuning of initial conditions and model parameters, the late time transition from matter domination towards an accelerated brane gravity dominated solution is obtained. An example trajectory which describes the evolution of the Universe from a stiff matter like solution (points $B_{2\pm}$) to a matter dominated epoch (point $B_1$) and then eventually towards an accelerated brane gravity solution $B_4$ is shown numerically in Fig.\ \ref{fig:para_w_eff_sca}. As in the case of the absence of the scalar field, Fig.\ \ref{fig:w_eff_sca} shows that $w_{\rm DE}$ diverges  but the overall effective equation of state ($w_{\rm eff}$) remains finite. Moreover, in Fig. \ref{fig:w_eff_sca} it can be seen that the effective DE eventually evolves as phantom DE in future time ($w_{\rm DE}<-1$) without converging to the cosmological constant value, unlike in the case of the absence of the scalar field. We can conclude that the effective behavior of this braneworld model mimics the cosmological dynamics of the $\Lambda$CDM model in the future time, even though it exhibits a phantom DE component, there is no danger of a Big-Rip singularity, since the overall $w_{\rm eff}$ converges to $-1$.

\subsection{Hyperbolic potential $V=V_0 \cosh ^{-\mu}(\lambda\phi)$}\label{subsec:cosh}
In this case, we consider the potential $V=V_0 \cosh ^{-\mu}(\lambda\phi)$ where $V_0$, $\lambda$ are constants of suitable dimension and $\mu$ is a dimensionless constant. This potential belongs to an $\alpha$-attractor family of potential which can lead to a DE tracker model with a late time cosmological constant behavior from a wide range of initial conditions \cite{Mishra:2017ehw, Bag:2017vjp}. Moreover, this potential has been also introduced to describe the evolution of the universe from scaling to de Sitter like attractor through spontaneous breaking mechanism \cite{Zhou:2007xp}. For this potential, we have
\begin{align}
f(s)=\frac{s^2}{\mu}-\mu \lambda^2,
\end{align}
hence
\begin{align}
s_*=\pm \mu \lambda~~~ \text{and}~~~~ df(s_*)=\frac{2s_*}{\mu}=\pm 2 \lambda.
\end{align}
Since the critical point $B_1$ is independent of the choice of the potential, its behavior is same as in the case of the general potential. However, each of the critical points $B_{2+}$, $B_{2-}$, $B_3$ and $B_5$ appear with two copies associated to two solutions $s_*=\pm \mu \lambda$. For this we shall use the notation $B_j^{+}$ for points correspond to $s_*=\mu\lambda$ and $B_j^{-}$ for $s_*=-\mu\lambda$.

The critical points $B_{2\pm}^+$, $B_{2\pm}^-$ correspond to stiff matter like solutions dominated by the kinetic part of the scalar field. Hence, none of these points correspond to accelerated universe. These points exist for any value of parameters $\mu$ and $\lambda$. The point $B_{2-}^+$ is unstable node when $\mu \lambda>\sqrt{6}$ and $\lambda<0$; the point $B_{2-}^-$ is unstable node when $\mu \lambda<-\sqrt{6}$ and $\lambda>0$ otherwise they are saddle. Point $B_{2+}^+$ is unstable node when $\mu \lambda>-\sqrt{6}$ and $\lambda>0$; point $B_{2+}^-$ is unstable node when $\mu \lambda<\sqrt{6}$ and $\lambda<0$, otherwise they are saddles. The scalar field dominated critical points $B_3^{\pm}$ exist when $\mu^2 \lambda^2<6$ but correspond to an accelerated universe when $\mu^2 \lambda^2<2$. However, both points are saddles in nature for any choice of the parameters $\mu$, $\lambda$ and $w$, hence they cannot describe late time universe. The stability of a nonhyperbolic set of critical points $B_4$ depends on the value of potential variable $s$ (see appendix \ref{app:1}). Since for this potential $f(s)=0$ for some $s$ i.e.\ for $s=\pm\mu \lambda$. Hence, points $(x_c,\sqrt{2 x_c+1},0,0,s_c)$ on the set $B_4$ are behaving as a late time attractor for $s_c=\pm \mu \lambda$ with $\mu>0$  and $\frac{w}{x_c+1} \geq 0$. This is important as it depicts the role of potential parameters $\mu$ and $\lambda$ complementing and influencing the brane gravity effects on the late time universe. The critical points $B_{5}^{\pm}$ correspond to unaccelerated scaling solutions but behave as if they correspond to matter dominated universe ($w_{\rm eff}=w$). Both these points exist for $\mu^2 \lambda^2<3$ but they are saddles for any choice of parameters $\mu$, $\lambda$ and $w$ (already discussed in the general potential case).

\begin{figure}
    \centering
    \subfigure[]{%
        \includegraphics[width=6cm,height=5cm]{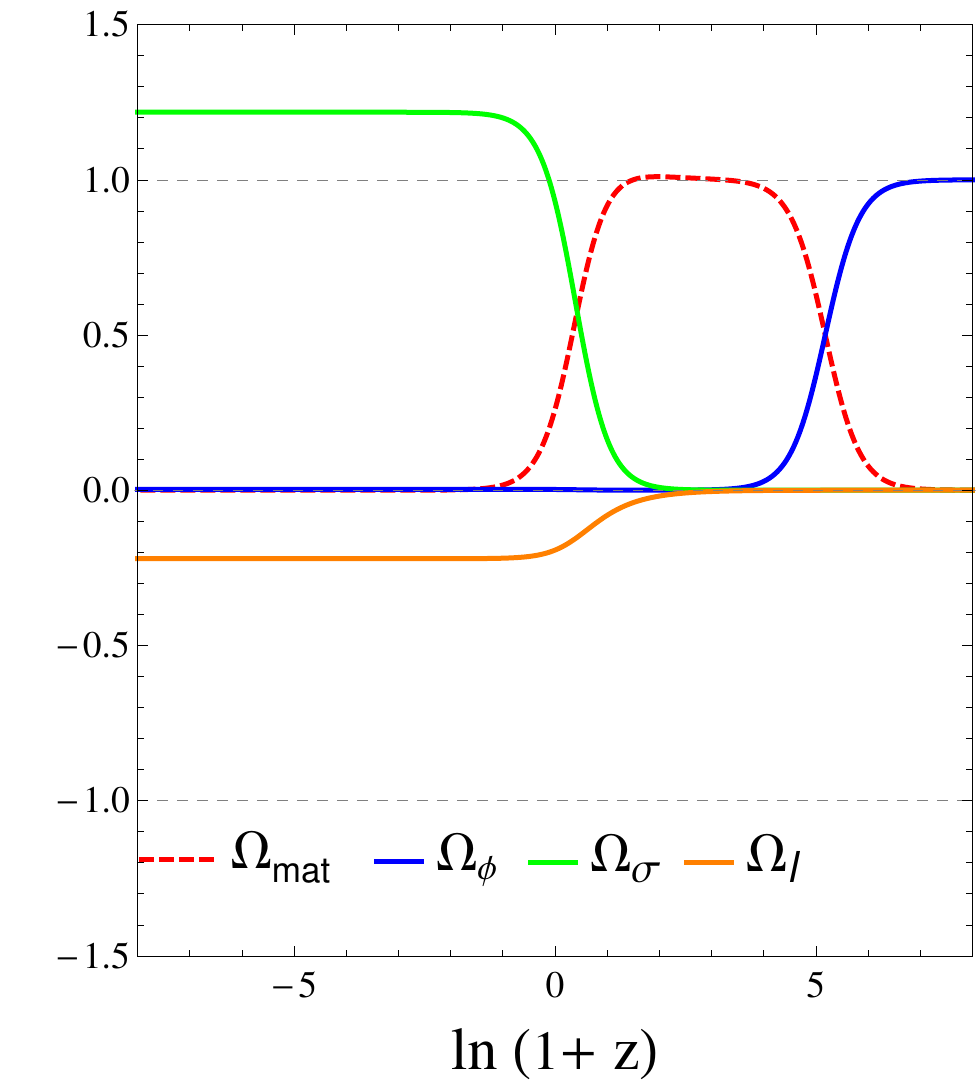}\label{fig:cosh_potential_density}}
    \qquad
    \subfigure[]{%
        \includegraphics[width=6cm,height=5cm]{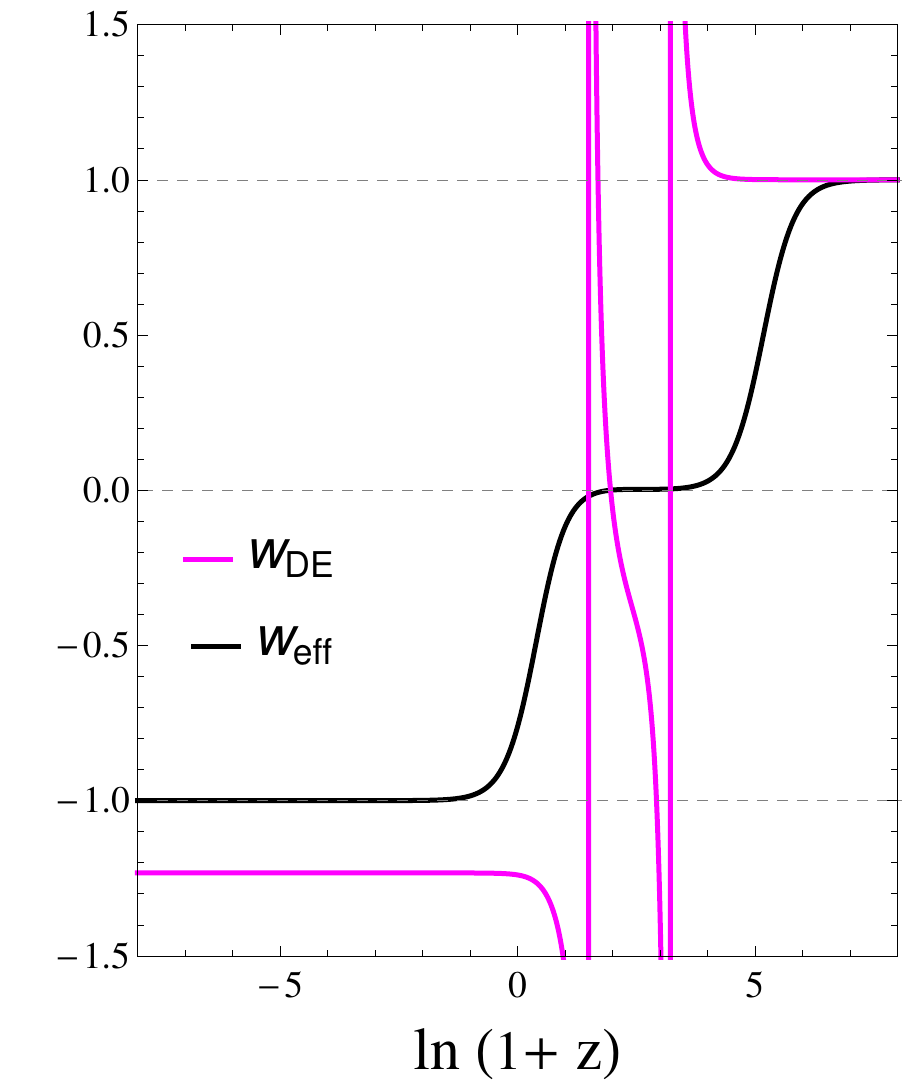}\label{fig:cosh_potential_weff}}
    \caption{(a) The time evolution of the relative matter energy density $\Omega_{\rm mat}$, the relative DE density $\Omega_\phi$ and the relative energy density due to brane corrections $\Omega_\sigma$ and $\Omega_l$. (b) The time evolution of the overall effective EoS $w_{\rm eff}$ and the effective EoS of DE $w_{\rm DE}$. In both panels, we take the potential $V=V_0 \cosh ^{-\mu}(\lambda\phi)$ with $\mu=2$, $\lambda=1$ and $w=0$.}
    \label{fig:para_w_eff_cosh}
    \end{figure}

Fig. \ref{fig:para_w_eff_cosh} shows the effect of brane gravity correction towards late time universe by plotting the evolution of relevant cosmological parameters. One such trajectory depicting the evolution of the Universe starting from a stiff matter phase (points $B_{2\pm}$) passing through towards a long-lasting matter dominated point $B_1$ and then eventually settling to a brane gravity ruled critical set $B_4$ mimicking the cosmological constant ($w_{\rm eff}=-1$). Therefore, as in the case of an exponential potential, the effective behavior of this braneworld model mimics that of the $\Lambda$CDM model in the future time, even though it exhibits a phantom DE component.

% Further,  a transition from a matter domination to a late time DE  era is characterized by oscillations of the overall effective EoS $w_{\rm eff}$, which bounce the universe between quintessence and cosmological constant domination before eventually stabilizing to a deSitter solution.

\subsection{Inverse powerlaw potential $V=\frac{V_0}{\phi^n}$}\label{subsec:pow}
Finally, we consider the case of powerlaw potential $V=\frac{V_0}{\phi^n}$ where $V_0$ is a constant of suitable dimension and $n$ is a dimensionless constant. The inverse powerlaw potential is well known for its tracking property where the scalar field with this potential tracks as the background dominant counterparts at any given cosmological time and ultimately dominates to lead to late time acceleration \cite{Zlatev:1998tr}. This potential has been also considered for inflation \cite{Barrow:1990vx}. For this potential, the function $f(s)=\frac{s^2}{n}$ (i.e. $\Gamma \neq 1$) and hence $s_*=0$.

In this case, the point $B_5$ does not exist. Since the critical point $B_1$ is independent of the choice of potential, its behavior remains the same as in the case of the general potential. Stiff matter dominated critical points $B_{2\pm}$  always behave as past time attractors. For this potential, the scalar field dominated point $B_3$ always corresponds to an accelerated solution for any choice of the model parameters. Unlike the potential where $s_* \neq 0$, this point is nonhyperbolic in nature. The detailed analysis of the center manifold theory is given in the appendix \ref{app:2}. From the analysis, it shows that this point is a saddle in nature. As we have seen in the general potential case, the nonisolated critical set $B_4$ arising from the brane gravity effects is also nonhyperbolic. For this potential, $f(s_c) \neq 0$ for nonzero $s_c$, hence the set $B_4$ is then behaving as a saddle. However, for $s_c=0$, the center manifold theory cannot extract the exact stability behavior of this set.

\begin{figure}
    \centering
    \subfigure[]{%
        \includegraphics[width=6cm,height=5cm]{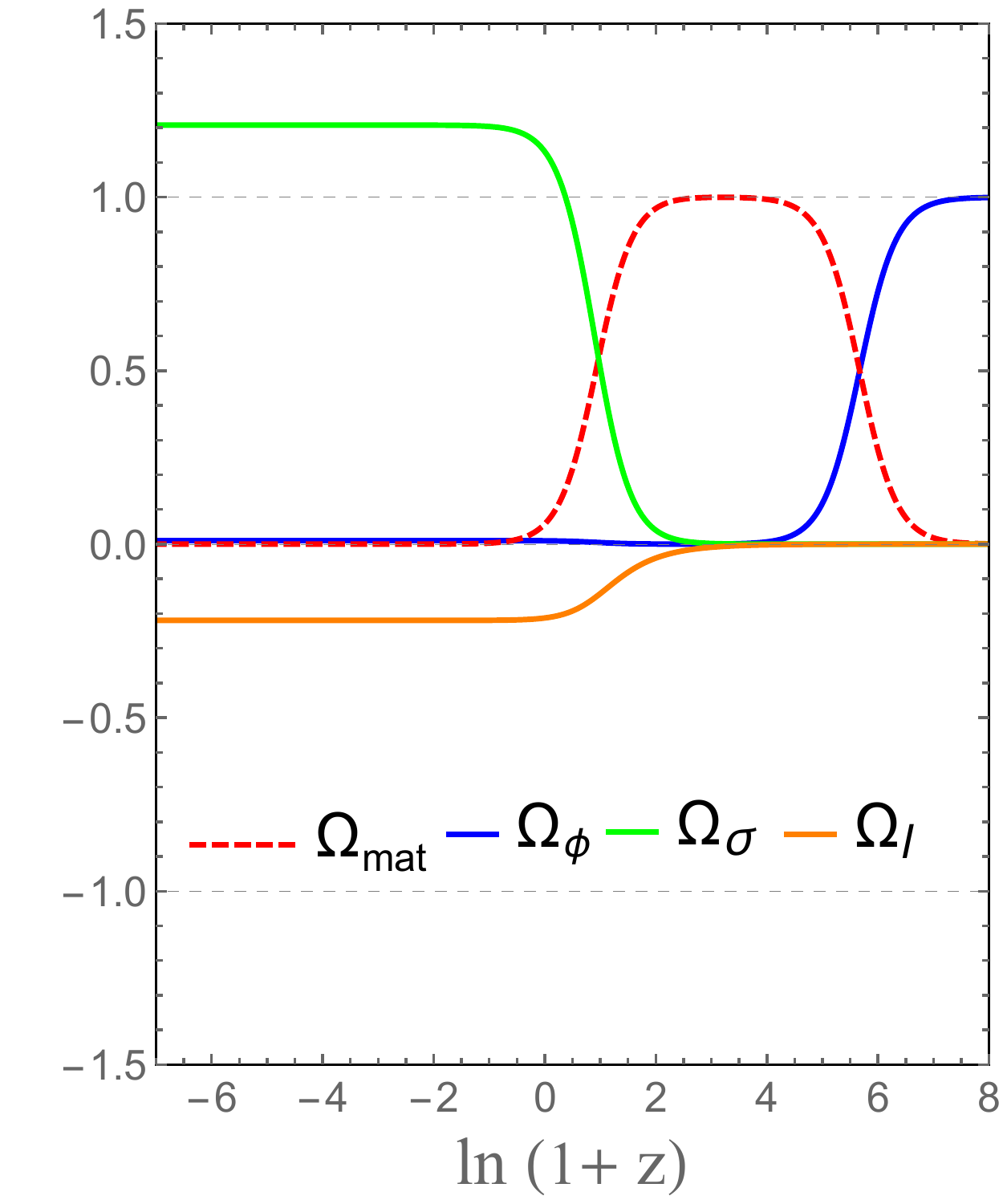}\label{fig:omega_powerlaw}}
    \qquad
    \subfigure[]{%
        \includegraphics[width=6cm,height=5cm]{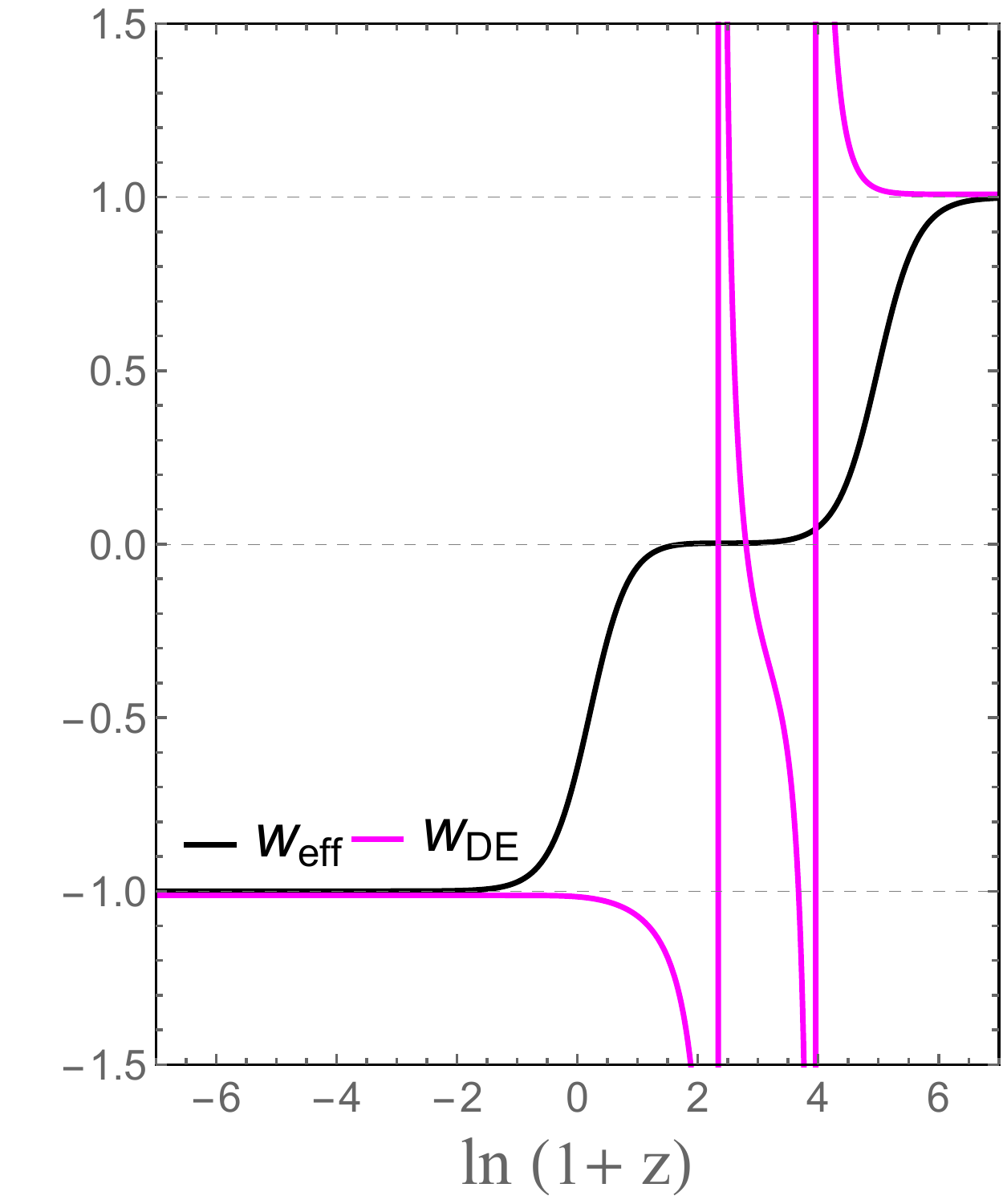}\label{fig:weff_powerlaw}}
    \caption{(a) The time evolution of the relative matter energy density $\Omega_{\rm mat}$, the relative DE density $\Omega_\phi$ and the relative energy density due to brane corrections $\Omega_\sigma$ and $\Omega_l$. (b) The time evolution of the overall effective EoS $w_{\rm eff}$ and the effective EoS of DE $w_{\rm DE}$. In both panels, we take $w=0 $ and the potential $V=\frac{V_0}{\phi^n}$ with $n=4$.}
    \label{fig:para_w_eff_powerlaw}
    \end{figure}

From the above analysis, we see that there is no generic late time attractor. Fig.\ \ref{fig:para_w_eff_powerlaw} depicts the evolution of the universe from a stiff matter dominated era $B_{2\pm}$ ($\Omega_\phi=1,\, w_{\rm eff}=1$) towards  an accelerating era dominated by the energy density due to the brane effects $B_4$ ($\Omega_\sigma+\Omega_l=1$, $w_{\rm eff}=-1$) through an intermediate era of matter domination $B_1$ ($\Omega_{\rm mat}=1$, $w_{\rm eff}=w$). Note that as the accelerated brane dominated phase is represented by a saddle set of critical points, the Universe will remain in this phase for a finite period of time.

In summary, the analysis performed above for three different potentials gives rise to contrasting results in late time universe for a physically interesting case of dust matter, $w=0$. For the exponential potential, the cosmological constant behavior due to brane gravity correction given by $B_4(x_c,\sqrt{2x_c+1},0,0,s_c)$ describes the late universe independent of initial conditions and choice of the model parameters. For the hyperbolic potential, the same behavior appears at the late time but with a specific choice of model parameters (i.e. $\mu>0$).  However, in the case of inverse power-law potential, the late time behavior of the Universe cannot be successfully explained even though the Universe evolves towards a de Sitter phase with brane gravity correction for a finite period of time.

Further, in each potential case, it can be verified that there is a possibility for $\Omega_{\rm mat}>1$ during an early matter-dominated era, similarly to the cases without an extra scalar field. Hence, the unusual property that the relative matter density can fluctuate above the critical value during the matter domination period is a generic phenomenon in the brane gravity model.

\section{Cosmological Implications} \label{sec:cos_imp}
In this section, we try to summarize the cosmological behavior of the braneworld model analyzed above. The cosmological dynamics of the braneworld model contains some interesting solutions with viable phenomenology such as matter domination, the effective phantom behavior of DE, cosmic mimicry, late time brane gravity effect. In what follows, we shall discuss each aspect separately:

\begin{itemize}
\item \textit{Matter dominated era}: This cosmic era corresponds to an unaccelerated intermediate stage of the Universe. It is characterized by
a long-lasting matter-dominated phase ($\Omega_{\rm mat}=1$) with an
effective EoS $w_{\rm eff}=w$. In this braneworld model, the matter
dominated universe is represented by a critical point $A_1$ which
behaves as an unstable node and  $B_1$ which is saddle in nature.
The matter dominated era is an important cosmic era as it is
responsible for large-scale structure formation. In both models, a
long-lasting matter dominated phase is smoothly followed by an
accelerated DE dominated phase which is in agreement with the
observed cosmological dynamics.

It is also observed that the model permits a bit unusual feature of relative matter overdensity, $\Omega_{\rm mat}>1$, during early matter domination era. This occurs when the effective energy density due to dark energy components, i.e. $\Omega_\sigma+\Omega_l$ or $\Omega_\sigma+\Omega_l+\Omega_{\phi}$ (in the presence of scalar field) is negative. This seems to be a generic phenomenon in brane gravity models, and it provides an important signature in view of astronomical observations.

\item \textit{Phantom behavior of DE}: The phantom behavior of DE corresponds to the solutions where $w_{\rm DE}<-1$, which is achieved only by a problematic phantom
 field in standard GR. Usually, in such a scenario, the fate of the Universe is a Big-Rip singularity at a finite time in the future. However, the braneworld overall
  effective matter description does not violate the weak energy condition, $p_{\rm eff}+\rho_{\rm eff} \geq 0$.
  This behavior can be seen from Figs. \ref{fig:weff_no_sca}, \ref{fig:w_eff_sca} where the effective EoS of DE encounters a singularity (i.e. $w_{\rm DE} \rightarrow \pm \infty$)
   but the effective EoS of the usual matter and brane gravity correction combined, $w_{\rm eff}$, remains finite and evolves towards $-1$. Hence, within the brane gravity scenario,
   a Big-Rip singularity is avoided even though the DE is phantomlike in the future time (see Fig. \ref{fig:w_eff_sca}).

\item \textit{Cosmic mimicry}: One of the important features of the braneworld model is that in future time the Universe expands as described by the $\Lambda$CDM model \cite{Bag:2018jle}. This property of the braneworld model is known as \textit{cosmic mimicry}. For example in the presence of a scalar field Fig.\ \ref{fig:para_w_eff_sca} shows a transition from a stiff matter dominated solution (points $B_{2\pm}$) towards an unaccelerated matter dominated solution (point $B_1$) and eventually to an accelerated phase due to the brane gravity effects (set $B_4$), mimicking a cosmological constant behavior at late times ($w_{\rm eff}=-1$). In the absence of a scalar field, however, the Universe evolves directly from a matter dominated phase towards  an accelerated phase due to the brane gravity effects (see Figs.\ \ref{fig:para_no_sca}, \ref{fig:weff_no_sca}). Therefore in all the cases, a smooth transition from matter domination towards an expanding de Sitter solution occurs and hence this braneworld model roughly evolves as the $\Lambda$CDM model.

\item \textit{Brane gravity effects}: The brane gravity effects are characterized by dynamical variables $x$ and $y$. The brane gravity effects are significant when the expression $y^2-2x$ is nonzero i.e.\ when $\Omega_{\sigma}+\Omega_l$ dominates the energy contribution (see Figs.  \ref{fig:para_no_sca}, \ref{fig:para_sca_1}). In both cases, the brane gravity corrections are represented by the critical sets $B_4$ or $A_2$. From the analysis, we see that both sets $A_2$ and $B_4$ are the only accelerated late time attractors mimicking the cosmological constant. Hence, the brane gravity plays a crucial role in the  late time universe. It may be mentioned that due to the brane gravity effects, the braneworld model behaves as a phantom model  and hence it is referred to as \textit{Phantom brane} (see Figs. \ref{fig:w_eff_sca}, \ref{fig:cosh_potential_weff}).

\item \textit{Observational signature}: The braneworld model analyzed in this paper resembles the behavior of the $\Lambda$CDM in the future times at the background level. This is illustrated numerically as in Figs. \ref{fig:weff_no_sca}, \ref{fig:w_eff_sca}, \ref{fig:cosh_potential_weff}  where the effective EoS evolves smoothly from $0$ to $-1$. However, as discussed above, the braneworld model possess a phantomlike behavior but without any Big-Rip singularity problem. It was also found that this braneworld model shows a significant deviation from the concordance model at the perturbation level \cite{Bag:2016tvc}. Hence, this scenario will indeed give distinct observational features with respect to the $\Lambda$CDM model and can thus in principle compared with astronomical observations.
\end{itemize}

\section{Conclusion}\label{sec:conc}
In this work, we have studied the cosmological dynamics of a general braneworld model in the presence of scalar field for a  wider class of potentials as well as in the absence of the scalar field by employing the dynamical system techniques. The stability analysis of critical points is handled by using the combination of linear stability analysis and center manifold theory. Moreover, in order to have a complete picture of the cosmological dynamics from the global perspective, we have extended our analysis of critical points near infinity. For this, we have employed the  Poincar\'e central projection method (Sec.\ \ref{sec:poincare}). However, critical points obtained in the infinite region are not of any cosmological interest.

In the absence of a scalar field, the dynamics of the model is very simple. It consists of one point $A_1$ which is behaving as a past time attractor and a set of critical points $A_2$ behaving as a late time attractor. For any choice of the parameters and initial conditions, the Universe starts evolving from a long-lasting matter dominated phase towards a late time accelerated phase due to  the brane gravity effects. Hence, the cosmological dynamics of this model mimics that of the $\Lambda$CDM  model at the background level.  It is also noted that the effective DE evolves as the phantom field but eventually converges to the cosmological constant value, $w_{\rm DE}=-1$.

In the presence of a scalar field, we have first studied the case for a wider class of potentials. The study for a wider class of potentials is well motivated by the low energy level of high energy physics. In the presence of a scalar field, for the case of a general potential (see Sec.\ \ref{sec:stability_pts}), we followed the approach by considering the quantity $\Gamma$ (cf. Eq. \eqref{Gamma}) as a function of variable $s$ (cf. Eq. \eqref{variable2}). However, to have a better picture of the dynamics we consider three potentials as examples belonging to three distinct categories: the exponential potential, the hyperbolic potential and  the inverse power-law potential.
 In all the cases, we obtained one additional critical set of points $B_4$ which is not present in the GR context. This actually implies the contribution of the brane gravity corrections towards late time acceleration. In all cases, we obtained a matter dominated critical point (point $B_2$) which is crucial for explaining the structure formation  at the background level.

For the exponential potential case (see Sec. \ref{subsec:exp}) with a broad choices of the model parameters and initial conditions, the universe starts evolving
 from a stiff matter dominated points $B_{2\pm}$ ($w_{\rm eff}=1$) towards the only late time brane gravity dominated solution (set $B_4$) with  $w_{\rm eff}=-1$
 through a long-lasting matter dominated point $B_1$  ($w_{\rm eff}=w$). This can also be seen numerically from Fig.\ \ref{fig:para_w_eff_sca}. For the hyperbolic potential,
  the cosmological evolution is similar to the case of exponential potential but unlike the case of the exponential potential, a fine-tuning of the model parameters are
   required to describe the late time behavior of the universe (see Fig.\ \ref{fig:cosh_potential_weff}). We also note that, in the case of the inverse power-law potential,
   the model cannot successfully describe the late time behavior of the Universe. Thus, the  scalar field potential plays a crucial role in determining the late time behavior of the universe. Moreover, in all cases, we see that in late times the effective DE behaves as a phantom DE but the evolution of the Universe mimics that of the  $\Lambda$CDM model and avoids the future Big-Rip singularity. This is a common behavior of the braneworld theory, hence this result from the dynamical system perspective supports the earlier work (see \cite{Sahni:2002dx,Lue:2004za,Alam:2005pb}).

In conclusion, we found that the braneworld model gives contrasting results with and without a scalar field for the early universe. It  also gives rise to differences
 in late time behavior in the presence of a scalar field for different potentials. While, in the absence of scalar field the universe evolves directly from matter domination,
  in the presence of scalar field the universe starts from stiff matter dominated epoch. Strictly speaking, a stiff solution in which the speed of sound equals
   the speed of light is not physically viable at the classical macroscopic level. However, this solution may be relevant only in the early universe as they play
   a vital role in the spectrum of relic gravity waves which stems during inflation \cite{Joyce:1997fc}.

On the other hand, it seems that the unusual behavior of the matter
energy density parameter, i.e. $\Omega_{\rm mat}>1$ during matter
domination era, is a generic phenomenon of this brane gravity model,
with or without an extra scalar field. This can imply interesting
observational signatures and perhaps could have something to do with
the puzzle of the EDGES signal for the 21 cm hyperfine transition
line of neutral hydrogen \cite{Bowman:2018yin}. However, the
discussion on the full implications of this phenomenon still remains
for further work.

Even though this model mimics the $\Lambda$CDM well at background level, it was shown that there is a possible
deviation at the perturbation level \cite{Bag:2016tvc}. This might lead to further interesting signatures for the present and future
 observations. It would, therefore, be worthwhile to further analyze the evolution of cosmological perturbations and the behavior of matter density perturbations by using dynamical system techniques. Such analysis will give a general conclusion over a wide range of initial conditions for perturbations. However, such analysis is beyond the scope of the present work and we leave it for the  future work as well.

\acknowledgments J.D. acknowledges the support of an Associate
program of IUCAA. L.J. was supported by the Estonian Ministry for
Education and Science through the Institutional Research Support
Project IUT02-27 and Startup Research Grant PUT790, as well as the
European Regional Development Fund through the Center of Excellence
TK133 ``The Dark Side of the Universe''. Finally, authors  would like to thank the anonymous reviewer for her/his constructive suggestions which lead to the improvement of the work.

\appendix
%\section{Appendix}
\section{Appendix}
\subsection{Center manifold dynamics for the point $B_4$ with general potential:} \label{app:1}
In this section, we will apply the center manifold theory to study the stability properties of the non-isolated set of critical points $B_4=(x_c,\sqrt{2x_c+1},0,0,s_c)$. The eigenvalues corresponding to these points are $-3$, $-3(1+w)$, $0$, $0$, $0$. The first step is to translate the point $(x_c,\sqrt{2x_c+1},0,0,s_c)$ to the origin by using the transformation: $x\rightarrow x+x_c$, $y\rightarrow y+\sqrt{2x_c+1}$, $u\rightarrow u$, $v\rightarrow v$ and $s\rightarrow s+s_c$. Under this transformation, the system of equations (\ref{y_p})-(\ref{s_p}) takes the form:\\
\begin{eqnarray}
    x'&=&\frac{3u^2(x+x_c)}{(1+x+x_c)}+\frac{3(x+x_c)(1+w)}{2(1+x+x_c)}\Bigg[1+2(x+x_c)-u^2-v^2-(y+\sqrt{2x_c+1})^2\Bigg]\label{a1} \\
    y'&=&\frac{3u^2(y+\sqrt{2x_c+1})}{(1+x+x_c)}+\frac{3(y+\sqrt{2x_c+1})(1+w)}{2(1+x+x_c)}\Bigg[1+2(x+x_c)-u^2-v^2-(y+\sqrt{2x_c+1})^2\Bigg]\\
u'&=&-3u+\sqrt{\frac{3}{2}}(s+s_c) v^2+u\Bigg[\frac{3u^2}{(1+x+x_c)}+\frac{3(1+w)}{2(1+x+x_c)}
    \Big(1+2(x+x_c)-u^2-v^2\nonumber\\&&-(y+\sqrt{2x_c+1})^2\Big)\Bigg]\\
v'&=&-\sqrt{\frac{3}{2}}(s+s_c) uv+v\Bigg[\frac{3u^2}{(1+x+x_c)}+\frac{3(1+w)}{2(1+x+x_c)}\Big(1+2(x+x_c)-u^2-v^2\nonumber\\&&-(y+\sqrt{2x_c+1})^2\Big)\Bigg]\\
    s'&=&-\sqrt{6}~u f(s+s_c). \label{a5}
\end{eqnarray}
The next step is to transform the system to Jordan form as
\begin{eqnarray}\label{a2}
        \frac{d \alpha}{dN}&=&A \alpha+f(\alpha,\beta), \\
        \frac{d \beta}{dN}&=&B \beta+g(\alpha,\beta),
    \end{eqnarray}
where $(\alpha,\beta)\in \mathbb{R}^c \times \mathbb{R}^s$ with $f$ and $g$ satisfying
\begin{equation}\label{a3}
            f(0,0)=0,~~~~~Df(0,0)=0,\qquad  g(0,0)=0,~~~~Dg(0,0)=0.
    \end{equation}
%\end{center}

Here $A$ is a $c\times c$ matrix corresponding to zero real part eigenvalues, $B$ is a $s \times s$ matrix corresponding to non-zero negative real part eigenvalues and $Df$ is the Jacobian matrix of $f$. In order to form a new basis of the Jacobian matrix of $B_4$, we introduce a new set of variables given by
\begin{center}
    $$  \left(
    \begin{array}{rcl}
    X\\
    Y\\
    U\\
    V\\
    S\\
    \end{array}
    \right)=\left(
    \begin{array}{ccccc}
    0&0&\frac{\sqrt{6}}{3}f(s_c)&0&0\\
    -\frac{\sqrt{2x_c+1}}{1+x_c}&\frac{2x_c+1}{1+x_c}&0&0&0\\
    0&0&-\frac{\sqrt{6}}{3}f(s_c)&0&1\\
    0&0&0&1&0\\
    \frac{\sqrt{2x_c+1}}{1+x_c}&-\frac{x_c}{1+x_c}&0&0&0\\
    \end{array}
    \right)~~\left(
    \begin{array}{rcl}
    x\\
    y\\
    u\\
    v\\
    s\\
    \end{array}
    \right)
    $$
\end{center}

Using the above transformation, the system of equation (\ref{a1})-\eqref{a5} can now be written as
\begin{center}
    $$  \left(
    \begin{array}{rcl}
    X'\\
    Y'\\
    U'\\
    V'\\
    S'\\
    \end{array}
    \right)=\left(
    \begin{array}{ccccc}
    -3&0&0&0\\
    0&-3(1+w)&0&0\\
    0&0&0&0\\
    0&0&0&0\\
    0&0&0&0\\
    \end{array}
    \right)
    \left(
    \begin{array}{rcl}
    X\\
    Y\\
    U\\
    V\\
    S\\
    \end{array}
    \right)+
    \left(
    \begin{array}{rcl}
    g_1\\
    g_2\\
    f_1\\
    f_2\\
    f_3\\
    \end{array}
    \right),
    $$
\end{center}
where $g_1, g_2, f_1$, $f_2$ and $f_3$ are polynomials of degree
greater than two in $(X, Y, U, V, S)$ with
\begin{eqnarray} \label{f1}
    f_1 (X, Y, U, V,S)&=&-\frac {1}{4 f \left( s_{{c}} \right) ^{2} \left( x_{{c}}\sqrt {2\,x_{{c}}+1}+2\,Sx_{{c}}+x_{{c}}Y+\sqrt {2\,x_{{c}}+1}+S \right)}\Big[4\, \left( f \left( s_{{c}} \right)  \right) ^{3}SU{V}^{2}+4\, \left(
    f \left( s_{{c}} \right)  \right) ^{3}S{V}^{2}X+\nonumber\\&&4\, \left( f \left( s_
    {{c}} \right)  \right) ^{3}S{V}^{2}s_{{c}}+12\,f \left( s_{{c}}
    \right) f \left( s_{{c}}+U+X \right) SX-24\,x_{{c}}XS \left( f
    \left( s_{{c}} \right)  \right) ^{2}-12\,w\,XY \left( f \left( s
    _{{c}} \right)  \right) ^{2}-24\,x_{{c}}X\nonumber\\&& Y \left( f \left( s_{{c}}
    \right)  \right) ^{2}+\sqrt {2\,x_{{c}}+1}\Big(9\,{X}^{3}-9\,w\,{X}^{3}-12\, \left( f \left( s_{{c}} \right)  \right) ^
    {2}X-12\,w\,XYS \left( f
    \left( s_{{c}} \right)  \right) ^{2}-6\,w\,X
    {Y}^{2} \left( f \left( s_{{c}} \right)  \right) ^{2}\nonumber\\&&-12\,XYS \left( f \left( s_{{c}} \right)  \right) ^{2}-6\,w\,X{S}^{2} \left( f \left( s_{{c}} \right)  \right)
    ^{2}-6\,w\,X{V}^{2} \left( f \left( s_{{c}}
    \right)  \right) ^{2}+4\, \left( f \left( s_{{c}} \right)  \right) ^{
        3}U{V}^{2}x_{{c}}+4\, \left( f \left( s_{{c}}
    \right)  \right) ^{3}{V}^{2}X\nonumber\\&&x_{{c}}+4\, \left( f
    \left( s_{{c}} \right)  \right) ^{3}{V}^{2}s_{{c}
    }x_{{c}}+12\,f \left( s_{{c}} \right) f \left( s_{{c}}+U+X \right)
    Xx_{{c}}+4\, \left( f \left( s_{{c}} \right)
    \right) ^{3}{V}^{2}X+4\, \left( f \left( s_{{c}}
    \right)  \right) ^{3}{V}^{2}s_{{c}}+12\nonumber\\&&\,f \left(
    s_{{c}} \right) f \left( s_{{c}}+U+X \right)X-6\,
    X{S}^{2} \left( f \left( s_{{c}} \right)  \right)
    ^{2}-6\,X{V}^{2} \left( f \left( s_{{c}} \right)
    \right) ^{2}-6\,X{Y}^{2} \left( f \left( s_{{c}}
    \right)  \right) ^{2}-12\, \left( f \left( s_{{c}} \right)  \right) ^
    {2}Xx_{{c}}\nonumber\\&&+4\, \left( f \left( s_{{c}} \right)
    \right) ^{3}U{V}^{2}\Big)+24\,f \left( s_{{c}}
    \right) f \left( s_{{c}}+U+X \right) SXx_{{c}}+12\,f \left( s_{{c}}
    \right) f \left( s_{{c}}+U+X \right) XYx_{{c}}+4\, \left( f \left( s_
    {{c}} \right)  \right) ^{3}\nonumber\\&&{V}^{2}Ys_{{c}}x_{{c}}-12\,w\,x_{{c}}X
    Y \left( f \left( s_{{c}} \right)  \right) ^{2}+4\, \left( f \left( s_
    {{c}} \right)  \right) ^{3}{V}^{2}XYx_{{c}}+8\, \left( f \left( s_{{c}
    } \right)  \right) ^{3}S{V}^{2}s_{{c}}x_{{c}}+4\, \left( f \left( s_{{
        c}} \right)  \right) ^{3}U{V}^{2}Yx_{{c}}\nonumber\\&&+8\, \left( f \left( s_{{c}}
\right)  \right) ^{3}SU{V}^{2}x_{{c}}+8\, \left( f \left( s_{{c}}
\right)  \right) ^{3}S{V}^{2}Xx_{{c}}-12\,XY \left( f \left( s_{{c}}
\right)  \right) ^{2}-12\,XS \left( f \left( s_{{c}} \right)
\right) ^{2}\Big]
\end{eqnarray}
\begin{eqnarray} \label{f2}
f_2(X, Y, U, V,S)&=&-\frac {3V}{4 f \left( s_{{c}} \right)^{2} \left( x_{{c}}\sqrt {2\,x_{{c}}+1}+2\,Sx_{{c}}+x_{{c}}Y+\sqrt {2\,x_{{c}}+1}+S \right)}\Big[2\,f \left( s_{{c}} \right) SUX+2\,f \left( s_{{c}} \right) SXs_{{c}}+
4\,w\,x_{{c}}Y\nonumber\\&&\left( f \left( s_{{c}} \right)  \right) ^{2}+4\,f
\left( s_{{c}} \right) S{X}^{2}x_{{c}}+2\,f \left( s_{{c}} \right) {X
}^{2}Yx_{{c}}+\sqrt {2\,x_{{c}}+1}\Big(2\,{V}^{2} \left( f \left( s_{{c}}
\right)  \right) ^{2}+3\,w\,{X}^{2}+2\,
{Y}^{2} \left( f \left( s_{{c}} \right)  \right) ^
{2}\nonumber\\&&+2\,f \left( s_{{c}} \right) {X}^{2}+2\,{S}^{2} \left( f \left( s_{{c}} \right)  \right) ^{2}-3\,
{X}^{2}+2\,f \left( s_{{c}}
\right) UX+2\,f \left( s_{{c}} \right) Xs_{{c}}+2
\,w\,{S}^{2} \left( f \left( s_{{c}} \right)
\right) ^{2}+2\,w\,{V}^{2} \left( f \left( s
_{{c}} \right)  \right) ^{2}\nonumber\\&&+2\,w\,{Y}^{2}
\left( f \left( s_{{c}} \right)  \right) ^{2}+4\,
YS \left( f \left( s_{{c}} \right)  \right) ^{2}+2\,f \left( s_{{c}} \right) {X}^{2}x_{{c}}+2\,f
\left( s_{{c}} \right) UXx_{{c}}+2\,f \left( s_{{
        c}} \right) Xs_{{c}}x_{{c}}\nonumber\\&&+4\,w\,YS \left( f
\left( s_{{c}} \right)  \right) ^{2}\Big)+4\,Y \left( f \left( s_{{c}}
\right)  \right) ^{2}+4\,f \left( s_{{c}} \right) SXs_{{c}}x_{{c}}+2
\,f \left( s_{{c}} \right) UXYx_{{c}}+2\,f \left( s_{{c}} \right) XYs_
{{c}}x_{{c}}\nonumber\\&&+4\,f \left( s_{{c}} \right) SUXx_{{c}}+4\,x_{{c}}Y
\left( f \left( s_{{c}} \right)  \right) ^{2}+4\,w\,Y \left( f
\left( s_{{c}} \right)  \right) ^{2}+2\,f \left( s_{{c}} \right) S{X}
^{2}\Big]\\
f_3(X, Y, U, V,S)&=&-\frac {3S}{4 f \left( s_{{c}} \right)^{2} \left( x_{{c}}\sqrt {2\,x_{{c}}+1}+2\,Sx_{{c}}+x_{{c}}Y+\sqrt {2\,x_{{c}}+1}+S \right)}\Big[2\sqrt {2\,x_{{c}}+1}\Big(2w\,{S}^{2} \left( f \left( s_{{c}}
\right)  \right) ^{2}+4\,w\,YS \left( f
\left( s_{{c}} \right)  \right) ^{2}\nonumber\\&&+2\,w\,{
    V}^{2} \left( f \left( s_{{c}} \right)  \right) ^{2}+2\,w\,{Y}^{2} \left( f \left( s_{{c}} \right)  \right) ^{2}+2
\,{S}^{2} \left( f \left( s_{{c}} \right)
\right) ^{2}+4\,YS \left( f \left( s_{{c}}
\right)  \right) ^{2}+2\,{V}^{2} \left( f \left(
s_{{c}} \right)  \right) ^{2}+2\,{Y}^{2} \left( f
\left( s_{{c}} \right)  \right) ^{2}+\nonumber\\&&3w X^2-3X^2\Big)+4\,w\,x_{{c}}Y \left( f
\left( s_{{c}} \right)  \right) ^{2}+4\,w\,Y \left( f \left( s_{{c}} \right)  \right) ^{2}+4\,x
_{{c}}Y \left( f \left( s_{{c}} \right)  \right) ^{2}+4\,Y \left( f \left( s_{{c}} \right)  \right) ^{2}\Big] \label{f3}
\end{eqnarray}

Following the center manifold theory, the coordinates which correspond to the non-zero eigenvalues $(X, Y)$ can be approximated in terms of $(U, V, S)$ by the functions
\begin{center}
    $$
    h(U, V, S)= \left(
    \begin{array}{ccc}
    ~~h_1(U, V, S) \\\\
    h_2(U, V, S)\\
    \end{array}
    \right)=
    \left(
    \begin{array}{ccc}
    a_{1}\,{S}^{2}+a_{2}\,{U}^{2}+a_{3}\,{V}^{2}+a_{4}\,SU+a_{5}\,SV+a_{6}
    \,UV+a_{7}\,{S}^{3}+a_{8}\,{U}^{3}+a_{9}\,{V}^{3}+a_{10}\,{S}^{2}U\\+a_{
        11}\,{S}^{2}V+a_{12}\,{U}^{2}V+a_{13}\,{U}^{2}S+a_{14}\,{V}^{2}S+a_{15
    }\,{V}^{2}U+a_{16}\,SUV+\mathcal{O}(U^4, V^4) \\\\
    b_{1}\,{S}^{2}+b_{2}\,{U}^{2}+b_{3}\,{V}^{2}+b_{4}\,SU+b_{5}\,SV+b_{6}
    \,UV+b_{7}\,{S}^{3}+b_{8}\,{U}^{3}+b_{9}\,{V}^{3}+b_{10}\,{S}^{2}U\\+b_{
        11}\,{S}^{2}V+b_{12}\,{U}^{2}V+b_{13}\,{U}^{2}S+b_{14}\,{V}^{2}S+b_{15
    }\,{V}^{2}U+b_{16}\,SUV+\mathcal{O}(U^4, V^4)\\
    \end{array}
    \right).
    $$
\end{center}
The quasilinear partial differential equation for the center manifold is given by\\
\begin{equation}\label{a7}
Dh(U, V, S)~[A(U, V, S)+f(U, V, S,~ h(U, V, S))]-Bh(U, V, S)-g(U, V, S,~ h(U, V, S))=0,
\end{equation}
where
\begin{center}
    $g(U, V, S)=    \left(
    \begin{array}{c}
    g_1(U, V, S)\\
    g_2(U, V, S)\\

    \end{array}
    \right)$,
    ~~~$f(U, V, S)= \left(
    \begin{array}{c}
    f_1(U, V, S)\\

    f_2(U, V, S)\\
    f_3(U, V, S)\\

    \end{array}
    \right)$,~~~
    $B=\left(
    \begin{array}{cc}
    -3&~~0\\
    ~~0&-3(1+w)\\

    \end{array}
    \right)$ and $A=\left(
    \begin{array}{ccc}
    0&0&0\\
    0&0&0\\
    0&0&0\\

    \end{array}
    \right)$.
\end{center}
In order to solve Eq. (\ref{a7}), we substitute the values of  $A$, $B$, $f$, $h$ and $g$ into it and we compare equal powers of $(U, V, S)$ to obtain the approximation of $h(U, V, S)$. After comparing equal powers of $U, V, S$ from both sides of Eq. (\ref{a7}), the only non zero constants are:
\begin{gather}
b_{1,3}=-\frac{1}{2}{\frac {\sqrt {2\,x_{{c}}+1}}{(x_{{c}}+1)(1+w)}},\quad a_3=\frac{s_c}{3}f \left( s_{{c}} \right),\quad b_{7}=\frac{1}{2}{\frac {2\,x_{{c}}+1}{ \left( x_{{c}}+1 \right) ^{2}(1+w)}},\quad b_{14}=\frac{1}{2}\frac {2\,x_{{c}}+1}{(1+w)(1+x_c)^2},\quad a_{15}=\frac{1}{3}f(s_c). \nonumber
\end{gather}

%\qquad ~b_1=0, \qquad ~a_2=0, \qquad ~b_2=0, \qquad ~a_3=0,\qquad ~b_3=0,\qquad \qquad ~a_4=0,\qquad ~b_4=0, \nonumber \\  ~a_5=0,\qquad ~b_5=0,  \qquad ~b_6=0,\qquad ~a_7=0, \qquad ~b_7=0

The dynamics near the center manifold is determined by the equations
\begin{equation}\label{a8}
U'=A(U, V, S)+f_1(U, V, S,~ h(U, V, S)),
\end{equation}
\begin{equation}\label{a9}
V'=A(U, V, S)+f_2(U, V, S,~ h(U, V, S)),
\end{equation}
\begin{equation}\label{a10}
S'=A(U, V, S)+f_3(U, V,S,~ h(U, V,S)),
\end{equation}
i.e.
\begin{eqnarray}
U'&=&-s_cf(s_c)\,V^2-\left[s_c\, df(s_c)+f(s_c)\right]U V^2+\mathcal{O}(4), \label{a11}\\
V'&=&-\frac{V^3}{2}\Bigg[\frac{(3w+ (w+1)s_c^2(x_c+1)) }{ \left( 1+x_{c}\right)  \left( 1+w \right) }\Bigg]-\frac{3}{2}\frac{w}{(1+x_c)(1+w)}\,S^2\,V+\mathcal{O}(4),  \label{a12} \\
S'&=&-\frac{3}{2}\Bigg[\frac{w}{ \left( 1+x_{c} \right)  \left( 1+w \right) }\Bigg]S^3-\frac{3}{2}\Bigg[\frac {w}{ \left( 1+x_{{c}} \right)  \left( 1+w \right) }\Bigg]S V^2+\mathcal{O}(4). \label{a13}
\end{eqnarray}

%As expected, the second order of Eqs. \eqref{a12} and \eqref{a13} vanishes and hence the dynamics near the center manifold upto second order are determined by $U'$. This actually happens since the two dimensional eigenspace spanned by eigenvector corresponds to two of the vanishing eigenvalue determines the direction of the tangent plane at each point of the critical set. Actually, we have also checked that this happens if we further increase the order of approximation.

In what follows we analyze the dynamics near the center manifold determine by equations (\ref{a11})-(\ref{a13}) for the case of $w=0$ and $w\neq 0$ separately:

\subsubsection{Case I: $w=0$}

\begin{itemize}
\item If $f(s_c)=0$ for all $s_c$ i.e. $f=0$, then the variable $s$ is constant (i.e. exponential potential). So, the following two dimensional system will determine the dynamics
\begin{eqnarray}
V'&=&-\frac{1}{2}\,{s_c^2}V^3+\mathcal{O}(4), \label{a11_0}\\
S'&=&\mathcal{O}(4). \label{a12_0}
\end{eqnarray}
In such a case, the set $B_4$ behaves as a stable set.
% We also noted that the third order of approximation of $U'$ vanishes. This is also as expected since the one dimensional eigenspace spanned by eigenvector corresponds to one of the vanishing eigenvalue determines the direction of the tangent  at each point of a one dimensional critical set $B_4$.

\item If $f(s_c)\neq 0$ for any nonzero $s_c$ then the second order approximation will determine the nature of stability. The set $B_4$ is behaving as a saddle for this case. However, if $s_c=0$, then it is stable if $f(0)>0$.

\item If $f(s_c)=0$ for some $s_c$ the set $B_4$ will be stable if $s_c\,df(s_c)>0$, it is behaving as a saddle if $s_c\,df(s_c)<0$.
\end{itemize}

\subsubsection{Case II: $ w \neq 0$}

\begin{itemize}
\item If $f(s_c)=0$ for all $s_c$ i.e. $f=0$, then the two dimensional system is given by
\begin{eqnarray}
V'&=&-\frac{V^3}{2}\Bigg[\frac{(3w+ (w+1)s_c^2(x_c+1)) }{ \left( 1+x_{c}\right)  \left( 1+w \right) }\Bigg]-\frac{3}{2}\frac{w}{(1+x_c)(1+w)}\,S^2\,V+\mathcal{O}(4),  \label{a11_1} \\
S'&=&-\frac{3}{2}\Bigg[\frac{w}{ \left( 1+x_{c} \right)  \left( 1+w \right) }\Bigg]S^3-\frac{3}{2}\Bigg[\frac {w}{ \left( 1+x_{{c}} \right)  \left( 1+w \right) }\Bigg]S V^2+\mathcal{O}(4). \label{a12_1}
\end{eqnarray}
 In such a case, the set $B_4$ behaves as a stable set if $\frac{w}{x_c+1}>0$.

 \item If $f(s_c)\neq 0$ for any nonzero $s_c$ then the second order approximation will determine the nature of stability. The set $B_4$ is behaving as a saddle for this case. However, if $s_c=0$, then it is stable if $f(0)>0$ and $\frac{w}{1+x_c}>0$.

\item If $f(s_c)=0$ for some $s_c$ the set $B_4$ will be stable if $s_c\,df(s_c)>0$  and $\frac{w}{1+x_c}>0$ and it is behaving as a saddle if $s_c\,df(s_c)<0$  or $\frac{w}{1+x_c}<0$.

\end{itemize}

%In particular, if $s_c=0$ then the set is stable if $f(0)>0$.

\subsection{Center manifold dynamics for the point $B_3$ with inverse powerlaw potential:} \label{app:2}
In this section we will apply the center manifold theory to study the stability properties of the critical point $B_3=(0,0,0,1,0)$. The eigenvalues corresponding to this points are $-3$, $-3(1+w)$, $0$, $0$. The first step is to translate the point $(0,0,0,1,0)$ to the origin by using the transformation: $x\rightarrow x$, $y\rightarrow y$, $u\rightarrow u$ and $v\rightarrow v+1$, $s \rightarrow s$. Under this transformation, the system of differential equations (\ref{y_p})-(\ref{s_p}) takes the form as
\begin{eqnarray}
x'&=& 3\,{\frac {{u}^{2}x}{1+x}}+3x(1+w)\Bigg[\frac { 1+2\,x-{u}^{2}- \left(
    v+1 \right) ^{2}-{y}^{2} }{2+2\,x}\Bigg], \label{a22} \\
y'&=& 3\,{\frac {{u}^{2}y}{1+x}}+3y(1+w)\Bigg[\frac { 1+2\,x-{u}^{2}- \left(
    v+1 \right) ^{2}-{y}^{2} }{2+2\,x}\Bigg],\\
u'&=&-3\,u+\frac{1}{2}\,\sqrt {6}~s \left( v+1 \right) ^{2}+u \Bigg[ 3\,{\frac {{u}^
        {2}}{1+x}}+3(1+w)\,{\frac {1+2\,x-{u}^{2}- \left( v+1 \right) ^{2}-{y}^{2}
    }{2+2\,x}} \Bigg],\\
v'&=&-\frac{1}{2}\,\sqrt {6}~u \left( v+1 \right) s+ \left( v+1 \right)  \Bigg[ 3\,{
    \frac {{u}^{2}}{1+x}}+3(1+w)\,{\frac {1+2\,x-{u}^{2}- \left( v+1 \right) ^{
            2}-{y}^{2}}{2+2\,x}} \Bigg],\\
s'&=&-{\frac {\sqrt {6}~u{s}^{2}}{n}}. \label{a26}
\end{eqnarray}

We now introduce a new set of variables given by
\begin{center}
    $$  \left(
    \begin{array}{rcl}
    X\\
    Y\\
    U\\
    V\\
    S\\
    \end{array}
    \right)=\left(
    \begin{array}{cccccc}
    -1&0&0&1&0\\
    0&0&1&0&-\frac{\sqrt{6}}{6}\\
    0&0&0&0&1\\
    1&0&0&0&0\\
    0&1&0&0&0\\
    \end{array}
    \right)~~\left(
    \begin{array}{rcl}
    x\\
    y\\
    u\\
    v\\
    s\\
    \end{array}
    \right).
    $$
\end{center}
Using the above transformation, the system of equation \eqref{a22}-\eqref{a26} can now be written as
\begin{center}
    $$  \left(
    \begin{array}{rcl}
    X'\\
    Y'\\
    U'\\
    V'\\
    S'\\
    \end{array}
    \right)=\left(
    \begin{array}{ccccc}
    -3&0&0&0&0\\
    0&-3(1+w)&0&0&0\\
    0&0&0&0&0\\
    0&0&0&0&0\\
    0&0&0&0&0\\
    \end{array}
    \right)
    \left(
    \begin{array}{rcl}
    X\\
    Y\\
    U\\
    V\\
    S\\
    \end{array}
    \right)+
    \left(
    \begin{array}{rcl}
    g_1\\
    g_2\\
    f_1\\
    f_2\\
    f_3\\
    \end{array}
    \right),
    $$
\end{center}
where $g_1, g_2, f_1$, $f_2$ and $f_3$  are polynomials of degree greater than two in $(X, Y, U, V, S)$ with
\begin{eqnarray}
f_1 (X, Y, U, V, S)&=&-\frac{1}{n}\Big[{U}^{3}+\sqrt {6}Y{U}^{2}\Big], \label{a14_1} \\
f_2 (X, Y, U, V, S)&=&-\frac{V}{4(1+V)} \Big[ 2\,w\,\sqrt {6}YU-2\,\sqrt {6}YU+6\,w\,{S}^{2}+ w\,{U}^{2}+6\,w\,{V}^{2}+12\,w\,XV+6\,w\,{X}^{2}\nonumber\\&&+6 \,w\,{Y}^{2}+6\,{S}^{2} -{U}^{2}+6\,{V}^{2}+12\,XV+6\,{X}^{2}+12\, w\,X-6\,{Y}^{2}+12\,X \Big],\label{a15}\\
 f_3 (X, Y, U, V, S)&=&-\frac{S}{4(1+V)}\Big[ 2\,w\,\sqrt {6}YU-2\,\sqrt {6}YU+6\,w\,{S}^{2}+w\,{U}^{2}+6\,w\,{V}^{2}+12\,w\,XV+6\,w\,{X}^{2}\nonumber\\&&+6\,w\,{Y}^{2}+6\,{S}^{2}-{U}^{2}+6\,{V}^{2}+12\,XV+6\,{X}^{2}+12\,w\,X-6\, {Y}^{2}+12\,X  \Big]. \label{a16}
\end{eqnarray}

Following the center manifold theory, the coordinates which correspond to the non-zero eigenvalues $(X, Y)$ can be approximated in terms of $(U, V, S)$ by the functions
\begin{center}
    $$
    h(U, V, S)= \left(
    \begin{array}{ccc}
    ~~h_1(U, V, S) \\\\
    h_2(U, V, S)\\
    \end{array}
    \right)=
    \left(
    \begin{array}{ccc}
    a_{1}\,{S}^{2}+a_{2}\,{U}^{2}+a_{3}\,{V}^{2}+a_{4}\,SU+a_{5}\,SV+a_{6}
    \,UV+a_{7}\,{S}^{3}+a_{8}\,{U}^{3}+a_{9}\,{V}^{3}+a_{10}\,{S}^{2}U\\+a_{
        11}\,{S}^{2}V+a_{12}\,{U}^{2}V+a_{13}\,{U}^{2}S+a_{14}\,{V}^{2}S+a_{15
    }\,{V}^{2}U+a_{16}\,SUV+\mathcal{O}(U^4, V^4) \\\\
    b_{1}\,{S}^{2}+b_{2}\,{U}^{2}+b_{3}\,{V}^{2}+b_{4}\,SU+b_{5}\,SV+b_{6}
    \,UV+b_{7}\,{S}^{3}+b_{8}\,{U}^{3}+b_{9}\,{V}^{3}+b_{10}\,{S}^{2}U\\+b_{
        11}\,{S}^{2}V+b_{12}\,{U}^{2}V+b_{13}\,{U}^{2}S+b_{14}\,{V}^{2}S+b_{15
    }\,{V}^{2}U+b_{16}\,SUV+\mathcal{O}(U^4, V^4)\\
    \end{array}
    \right).
    $$
\end{center}
The quasilinear partial differential equation for the center manifold is given by
\begin{equation}\label{a30}
Dh(U, V, S)~[A(U, V, S)+f(U, V, S,~ h(U, V, S))]-Bh(U, V, S)-g(U, V, S,~ h(U, V, S))=0,
\end{equation}
where
\begin{center}
    $g(U, V, S)=    \left(
    \begin{array}{c}
    g_1(U, V, S)\\
    g_2(U, V, S)\\

    \end{array}
    \right)$,
    ~~~$f(U, V, S)= \left(
    \begin{array}{c}
    f_1(U, V, S)\\

    f_2(U, V, S)\\
    f_3(U, V, S)\\

    \end{array}
    \right)$,~~~
    $B=\left(
    \begin{array}{cc}
    -3&~~0\\
    ~~0&-3(1+w)\\

    \end{array}
    \right)$and $A=\left(
    \begin{array}{ccc}
    0&0&0\\
    0&0&0\\
    0&0&0\\

    \end{array}
    \right)$.
\end{center}

In order to solve Eq. (\ref{a30}), we substitute the values of  $A$, $B$, $f$, $h$ and $g$ into it and we compare equal powers of $(U, V, S)$ to obtain the approximation of $h(U, V, S)$. After comparing equal powers of $U, V, S$ from both sides of Eq. (\ref{a30}), the only non zero coefficients of $h_1$, $h_2$ are:
\begin{gather}
a_{1,3}=-\frac{1}{2}, \quad ~a_2=-\frac{1}{12},\quad ~b_6=\frac{\sqrt{6}}{3},\quad ~b_8=\frac{1}{18}\,{\frac {\sqrt {6}}{n}},
\quad ~b_{10}=-\frac{\sqrt{6}}{6}, \quad ~a_{11}=1,\quad
~a_{12}=-\frac{1}{6},\nonumber\\ a_{15}=-\frac{\sqrt{6}}{24}(1+w)-\frac{w}{16}.\nonumber
\end{gather}
The dynamics near the center manifold is determined by the equations
\begin{equation}\label{a18}
U'=A(U, V, S)+f_1(U, V, S,~ h(U, V, S)),
\end{equation}
\begin{equation}\label{a19}
V'=A(U, V, S)+f_2(U, V, S,~ h(U, V, S)),
\end{equation}
\begin{equation}\label{a20}
S'=A(U, V, S)+f_3(U, V,S,~ h(U, V,S)),
\end{equation}
i.e.
\begin{equation}\label{a34}
U'=-{\frac {{U}^{3}}{n}}+\mathcal{O}(4),
\end{equation}
\begin{equation}\label{a35}
V'=\frac{1}{2}\,{U}^{2}V+\mathcal{O}(4),
\end{equation}
\begin{equation}\label{a36}
S'=\frac{1}{2}\,{U}^{2}S+\mathcal{O}(4).
\end{equation}
From these three equations (\ref{a34}), (\ref{a35}) and (\ref{a36}), we can conclude that the point $B_3$ is saddle for any choice of $n$.

\end{document}